\documentclass[12pt]{article}
\usepackage[sfdefault]{ClearSans} %% option 'sfdefault' activates Clear Sans as the default text font
\usepackage[T1]{fontenc}

\usepackage{natbib}

\input{glyphtounicode}  % With this ``fi'' and other ligatures are copied correctly from the PDF
\usepackage{appendix}
\usepackage{pdflscape}
\usepackage{xspace}	
\usepackage{graphicx}
\usepackage[svgnames]{xcolor}
\usepackage[most]{tcolorbox}
\usetikzlibrary{shadows}
\usepackage{authblk}
\newcounter{exa}

\tcbset{
myexample/.style={
  enhanced,breakable,
  colback=yellow!10!white,
  colframe=red!50!black,
  fonttitle=\scshape,
  titlerule=0pt,
  title={\refstepcounter{exa}MoReBikeS example~\theexa.},
  title style={fill=yellow!10!white},
  coltitle=red!50!black,
  drop shadow,
  highlight math style={reset,colback=LightBlue!50!white,colframe=Navy}
  }
}

\newtcolorbox{texample}{myexample}

\usepackage{color}
\definecolor{orange}{RGB}{255,127,0}
\definecolor{brown}{RGB}{150,70,0}
\definecolor{red}{RGB}{255,90,90}
\definecolor{darkred}{RGB}{150,0,0}
\definecolor{myred}{RGB}{200,50,50}
\definecolor{green}{RGB}{127,255,127}
\definecolor{darkgreen}{RGB}{0,127,0}
\definecolor{mygreen}{RGB}{60,180,60}
\definecolor{lightblue}{RGB}{150,150,255}
\definecolor{blue}{RGB}{127,127,255}
\definecolor{darkblue}{RGB}{0,0,127}
\definecolor{myblue}{RGB}{80,80,200}
\definecolor{grey}{RGB}{127,127,127}
\definecolor{pink}{RGB}{255,180,180}
\definecolor{lightgrey}{RGB}{180,180,180}

\usepackage{verbatim}
\usepackage{array}
\usepackage{hyperref}
\usepackage{graphicx}

\usepackage{amsmath}
\usepackage{amssymb}

\graphicspath{{}}

\begin{document}
\title{\includegraphics[scale=0.6]{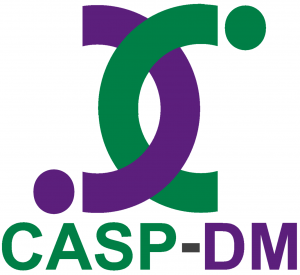} \\
Context Aware Standard Process for Data Mining\\
  \large \url{http://www.casp-dm.org}\vspace{1cm}}

\author[1]{\normalsize Fernando Mart{\mbox{\'{\i}}}nez-Plumed}
\author[1]{\normalsize Lidia Contreras-Ochando}
\author[1]{\normalsize C\`{e}sar Ferri}
\author[2]{\normalsize Peter Flach}
\author[1]{\normalsize Jos$\acute{\mbox{e}}$ Hern$\acute{\mbox{a}}$ndez-Orallo}
\author[3]{\normalsize Meelis Kull}
\author[4]{\normalsize Nicolas Lachiche}
\author[1]{\normalsize Mar${\mbox{\'{\i}}}$a Jos$\acute{\mbox{e}}$ Ram${\mbox{\'{\i}}}$rez-Quintana}

\affil[1]{\small Departament de Sistemes Inform\`atics i Computaci\'o, Universitat Polit\`ecnica de Val\`encia, Spain.
       \{\tt fmartinez,cferri,jorallo,mramirez\}@dsic.upv.es}
       
\affil[2]{\small University of Bristol, U.K. {\tt peter.flach@bristol.ac.uk}}
       
\affil[3]{\small University of Tartu, Estonia. \tt meelis.kull@ut.ee}

\affil[4]{\small ICube, Universit\'e de Strasbourg, France. \tt nicolas.lachiche@unistra.fr}

\renewcommand\Authands{ and }

%\author{\small Fernando Mart{\mbox{\'{\i}}}nez-Plumed, C\`{e}sar Ferri, Jos$\acute{\mbox{e}}$ Hern$\acute{\mbox{a}}$ndez-Orallo, Mar${\mbox{\'{\i}}}$a Jos$\acute{\mbox{e}}$ Ram${\mbox{\'{\i}}}$rez-Quintana\\
%\small DSIC, Universitat Polit\`ecnica de Val\`encia, Cam\'{\i} de Vera s/n, 46022 Val\`encia, Spain.\\
%\small E-mails: {\tt \{fmartinez,cferri,jorallo,mramirez\}@dsic.upv.es}}% <-this % stops a space
%\thanks{}% <-this % stops a space
%\thanks{}% <-this % stops a space
%\thanks{}}
%}
\date{}
\maketitle
\vspace{3cm}
\begin{abstract}

We propose an extension of the Cross Industry Standard Process for Data Mining (CRISP-DM) which addresses specific challenges of machine learning and data mining for context and model reuse handling. This new general context-aware process model is mapped with CRISP-DM reference model proposing some new or enhanced outputs.

\end{abstract}

\newpage

%
%KDD \citep{Fayyad1996KDD, fayyad1996advances}
%Human-centered \citep{Brachman1996PKD,gertosio2004knowledge}
%SEMMA \citep{SEMMA}
%5As \citep{martinez2003optimizacion}
%6-sigma \citep{phillips2005six}
%Cabena \citep{cabena1998discovering}
%Two Crows \citep{edelstein1998introduction,crows1999introduction}
%Anand and Buchner \citep{anand1998decision, anand1998data, buchner1999internet}
%CRISP-DM \citep{chapman2000crisp}
%KDD Roadmap \citep{debuse2001building}
%Marban \citep{Marban2007,marban2009toward}
%DMIE \citep{solarte2002proposed}
%RAMSYS \citep{moyle2001ramsys}
%Cios \citep{cios2000knowledge, cios2005trends}
\section{Introduction}

Anticipating potential changes in context is a critically important part of data mining projects.
Unforeseen context changes can lead to substantial additional costs and in the extreme case require running a new project from scratch.
For example, an automatic text summarisation system developed in the context of the English language can be extremely hard to be modified for other languages, unless such context change is anticipated.
For another example, a fraud detection service provider develops its detectors in the context of known types of frauds, but the context keeps changing, with new types invented continuously.
A careful analysis can help to build more versatile detectors which are effective for some new types of frauds and are easy to update for other new types.
As a third example, a customer segmentation system helping to tailor products for different customer groups might be hard to modify to incorporate richer customer information, unless such context changes are anticipated.

\begin{figure}[h]
	\centering
		\includegraphics[width=0.700\columnwidth]{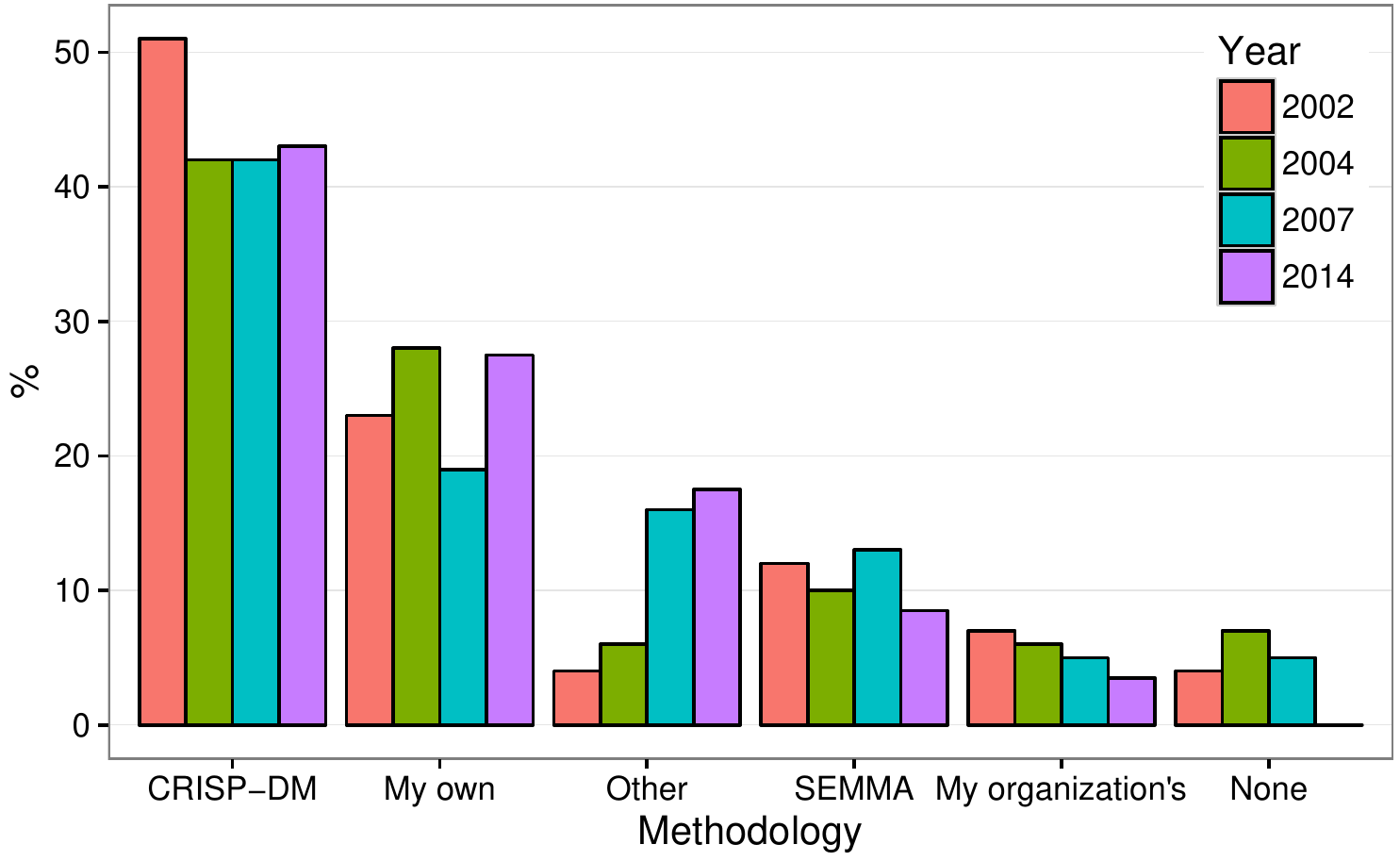}
	\caption{Use of data mining methodologies according to \url{www.kdnuggets.com}}
	\label{fig:trends}
\end{figure}

Context anticipation is more than just a single separate task and it requires dedicated activities in all phases of the data mining process, from the initial domain understanding up to the final deployment.
These activities are not included in any of the existing Data Mining (DM) standard process methodologies, such as the \emph{Knowledge Discovery in Databases} (KDD) Process \cite{Fayyad1996KDD}, the \emph{Cross Industry Standard Process for Data Mining} (CRISP-DM) \cite{chapman2000crisp} and the \emph{Sample, Explore, Modify, Model and Assess} (SEMMA)\cite{SEMMA} process model. In this paper, we report on an extension of the CRISP-DM process model called CASP-DM (\emph{Context-Aware Standard Process for Data Mining}), which has been evolving as a new standard with the goal of integrating context-awareness and context changes in the knowledge discovery process, while remaining backward compatible, so that users of CRISP-DM can adopt CASP-DM easily.

\begin{figure}[h]
	\centering
		\includegraphics[width=0.80\columnwidth]{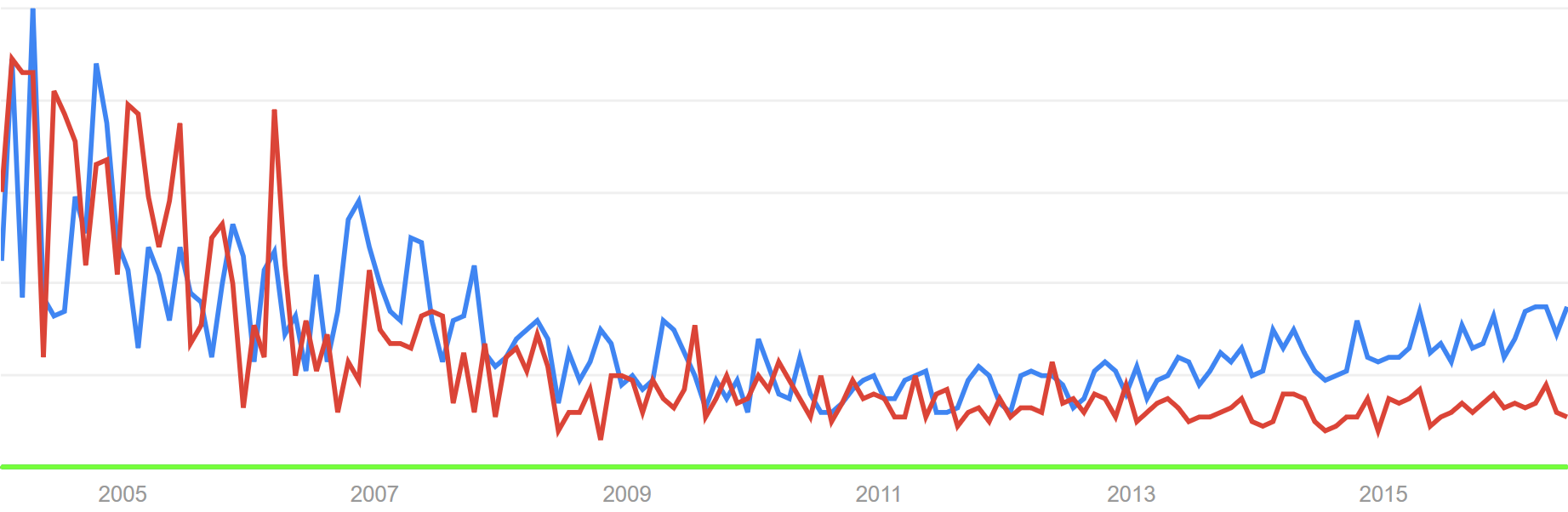}
	\caption{Relative interest over time in web searches according to Google Trends (\url{www.google.es/trends/}). Terms legend: CRISP-DM in \textcolor[rgb]{0,0,1}{blue}, KDD in \textcolor[rgb]{1,0,0}{red}, SEMMA in \textcolor[rgb]{0,0.58,0}{green} (the latter having a relative interest close to zero).}
	\label{fig:MethodologyUSE}
\end{figure}

The reasons why we have use CRISP-DM as a base are multiple. CRISP-DM %\cite{chapman2000crisp}, which was proposed in 2000 to 
is the most complete data mining methodology in terms of meeting the needs of industrial projects and has become the most widely used process for DM projects, according to the KDnuggets polls held in 2002, %\footnote{\url{http://www.kdnuggets.com/polls/2002/methodology.htm}}, 
2004, %\footnote{\url{http://www.kdnuggets.com/polls/2004/data\_mining\_methodology.htm}}, 
2007, %\footnote{http://www.kdnuggets.com/polls/2007/data\_mining\_methodology.htm} 
and 2014. %\footnote{\url{http://www.kdnuggets.com/polls/2014/analytics-data-mining-data-science-methodology.html}}. 
Although CRISP-DM does not seem to be maintained\footnote{The original crisp-dm.org site is no longer active.} or adapted to the new challenges in data mining, the proposed six phases and their subphases are still a good guide for the knowledge discovery process. In fact, the interest in CRISP-DM continues to be high compared to other models (see Figures \ref{fig:trends} and \ref{fig:MethodologyUSE}). Therefore, the participation and cooperation of the data mining community is, of course, pivotal to the success of CASP-DM. This inclusion should imply the development of a platform where the data mining community can have access to the standard, which otherwise has the risk of being diluted,  while working as an embryo for a committee and stable working group for an evolving standard accommodating future challenges and evolution of the field. Furthermore, CRISP-DM is supported by several project management software tools, such as RapidMiner\footnote{\url{https://rapidminer.com}} and IBM SPSS Modeler\footnote{\url{http://www.ibm.com/software/analytics/spss/products/modeler}}. The extension of CRISP-DM into CASP-DM allows data mining projects to become context-aware while keep using these tools. 

The rest of the document is organised as follows. 
Section~\ref{ref:review} briefly reviews CRISP-DM and related methodologies, and the state of the art in terms of standardisation and maintenance of the methology. 
Section~\ref{sec:context} discusses the role that context (or domain) is having in DM applications and the main types of context and context changes (including changes in costs, data distribution and others). 
Section~\ref{sec:casposo} proposes CASP-DM, with new tasks and outputs as well as enhancements to the original reference model thus allowing  the practitioners to be aware  of (and anticipate) the main types of context. Finally, section~\ref{lab:discussion} closes the paper.

\section{Review of DM and CRISP-DM methodologies}\label{ref:review}

\begin{figure}[h]
	\centering
		\includegraphics[width=1.00\textwidth]{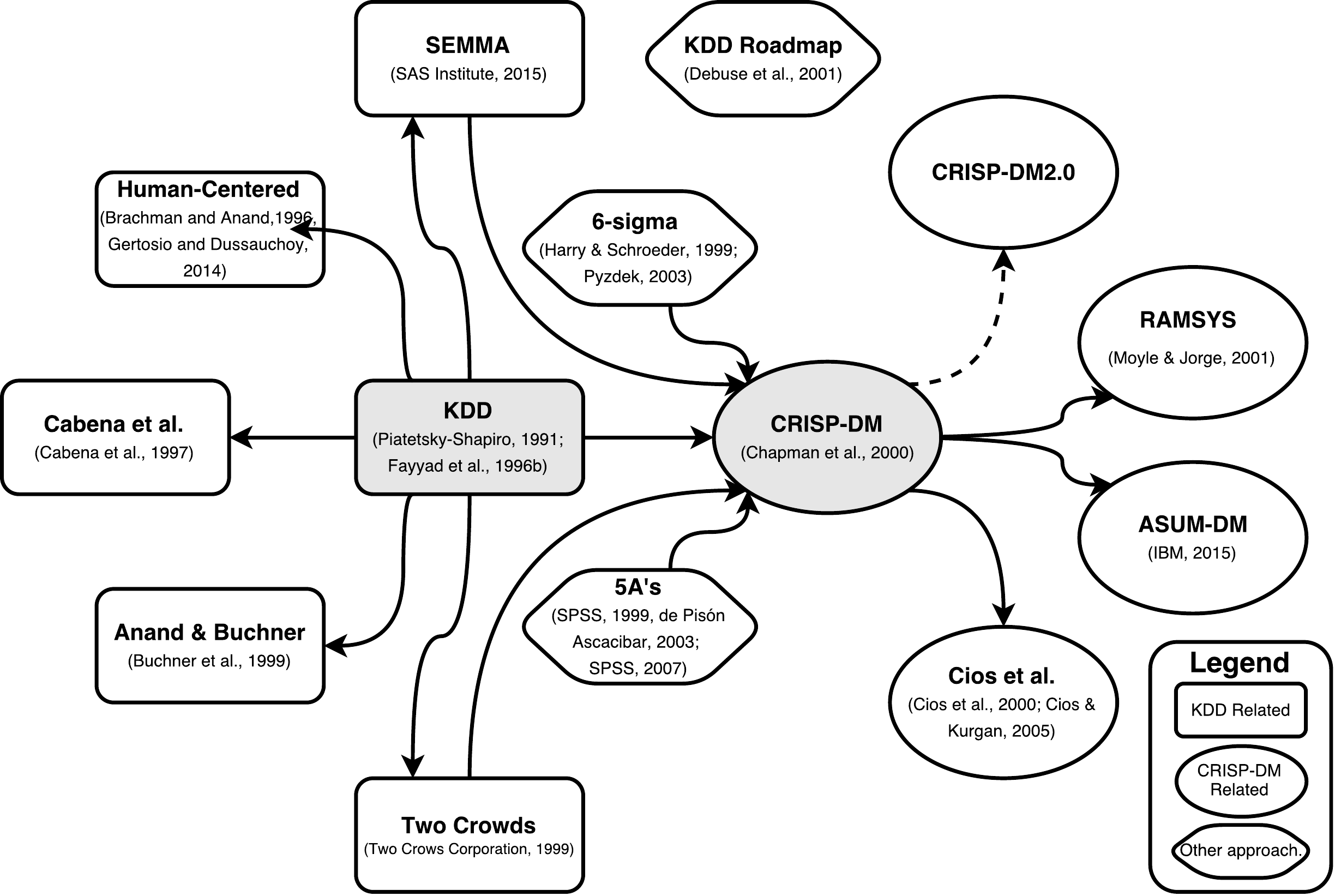}
	\caption{Evolution of DM Methodologies. Adapted from \citep{mariscal2010survey}}
	\label{fig:EvolutionKDD}
\end{figure}

\begin{figure}[h]
	\centering
		\includegraphics[width=1\textwidth]{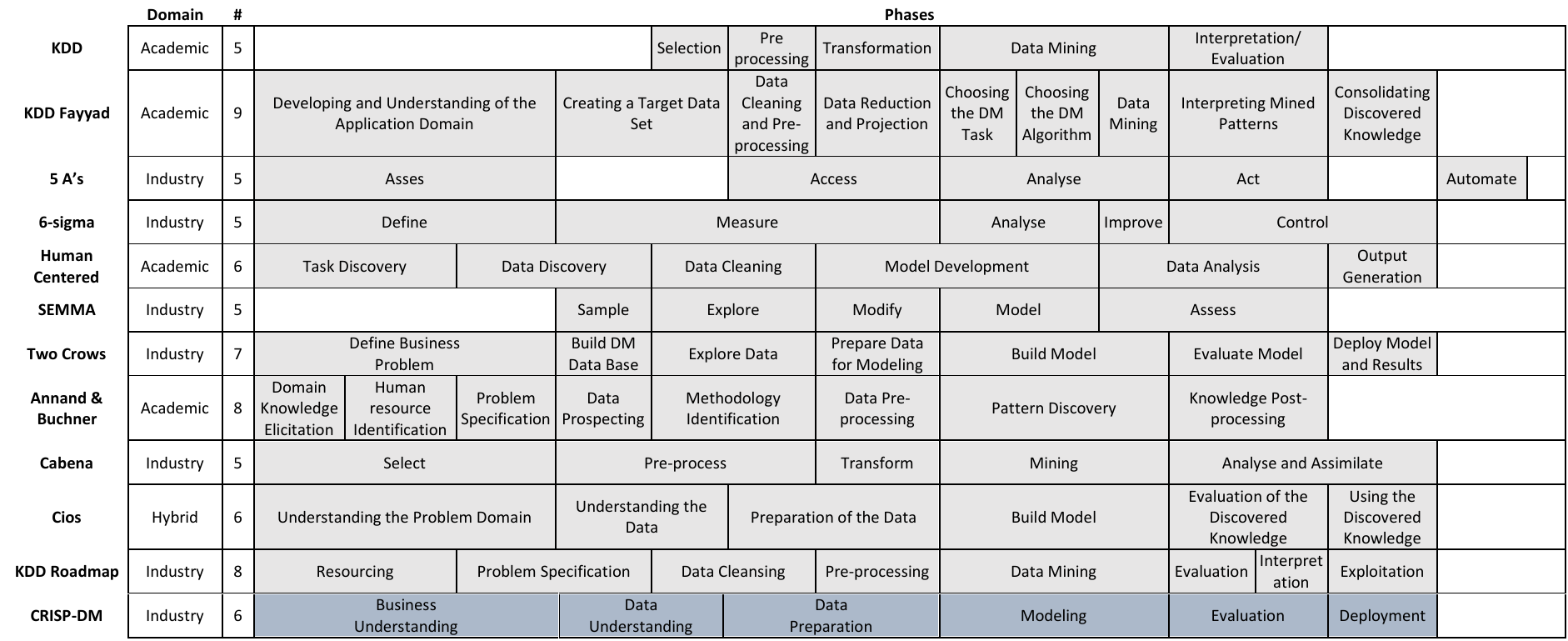}
	\caption{Phases of Data Mining  Methodologies}
	\label{fig:PhasesMethodologies}
\end{figure}

In this section we review the main approaches (process models and methodologies\footnote{While  a \emph{process model} is defined as a set of tasks to be performed to develop a particular element (as well as their inputs and outputs), a \emph{methodology} can be defined as a process model instance, in which not only tasks, inputs and outputs must be specified but also the way in which the tasks must be carried out.}) useful to extract useful information from large volumes of data (see \cite{mariscal2010survey} for a complete survey). We focus on two main approaches: \emph{Knowledge Discovery in Databases} (KDD) \citep{Fayyad1996KDD, fayyad1996advances} since it was the original approach, and the CRISP-DM \citep{chapman2000crisp}, since it is the reference methodology. The rest of approaches detailed are based on them. Figure \ref{fig:EvolutionKDD} shows a diagram of how the different DM and KD process models and methodologies have evolved. Furthermore, Table \ref{fig:PhasesMethodologies} compares the phases into which the DM and KD process is decomposed according the proposals discussed.

\subsection{KDD related approaches}

The term \emph{Knowledge Discovery in Databases} (KDD) \citep{Fayyad1996KDD, fayyad1996advances} was the first process model to establish all the steps to be taken to develop a Data Mining project. According to Fayyad et al. \cite{Fayyad1996KDD} KDD is defined as "[\dots] the non-trivial process of identifying valid, novel, potentially useful, and ultimately understandable patterns in data." The non-trivial\footnote{Involving search or inference.} goal is thus to (automatically) extract high-level knowledge that may not be obvious but potentially useful from raw or unprocessed data. This discovery of knowledge from a set of facts is accomplished by applying \emph{Data Mining} (DM) methods. However, KDD has a much broader scope, of which DM is just one step in the whole process model. This process model involves several steps, including data processing, search for patterns, knowledge evaluation and interpretation, and refinement, where the whole process is interactive and iterative, which  means  that sometimes it may be necessary to repeat the previous steps. The overall process involves the repeated application of the following nine steps:

\begin{itemize}
	\item \textbf{Developing an understanding of the application domain}, the relevant prior knowledge and the goals of the end-user.
  \item \textbf{Creating a target data set}: selecting a data set, or focusing on a subset of variables, or data samples, on which discovery is to be performed.
	\item \textbf{Data cleaning and preprocessing}: including basic operations for removing noise or outliers, collecting necessary information to model or account for noise, deciding on strategies for handling missing data fields, and accounting for time sequence information and known changes.
   \item \textbf{Data reduction and projection}: including finding useful features to represent the data depending on the goal of the task, using dimensionality reduction or transformation methods to reduce the effective number of variables under consideration or to find invariant representations for the data.
   \item \textbf{Choosing the data mining task}: deciding whether the goal of the KDD process is classification, regression, clustering, etc.
   \item \textbf{Choosing the data mining algorithm(s)}: selecting method(s) to be used for searching for patterns in the data, deciding which models and parameters may be appropriate and matching a particular data mining method with the overall criteria of the KDD process.
   \item \textbf{Data mining}: searching for patterns of interest in a particular representational form or a set of such representations as classification rules or trees, regression, clustering, and so forth.
   \item \textbf{Knowledge interpretation}: interpreting the discovered patterns.
   \item  \textbf{Consolidating discovered knowledge}: incorporating the discovered knowledge into the performance systems. 
\end{itemize}

\begin{figure}[h]
	\centering
		\includegraphics[width=1.00\textwidth]{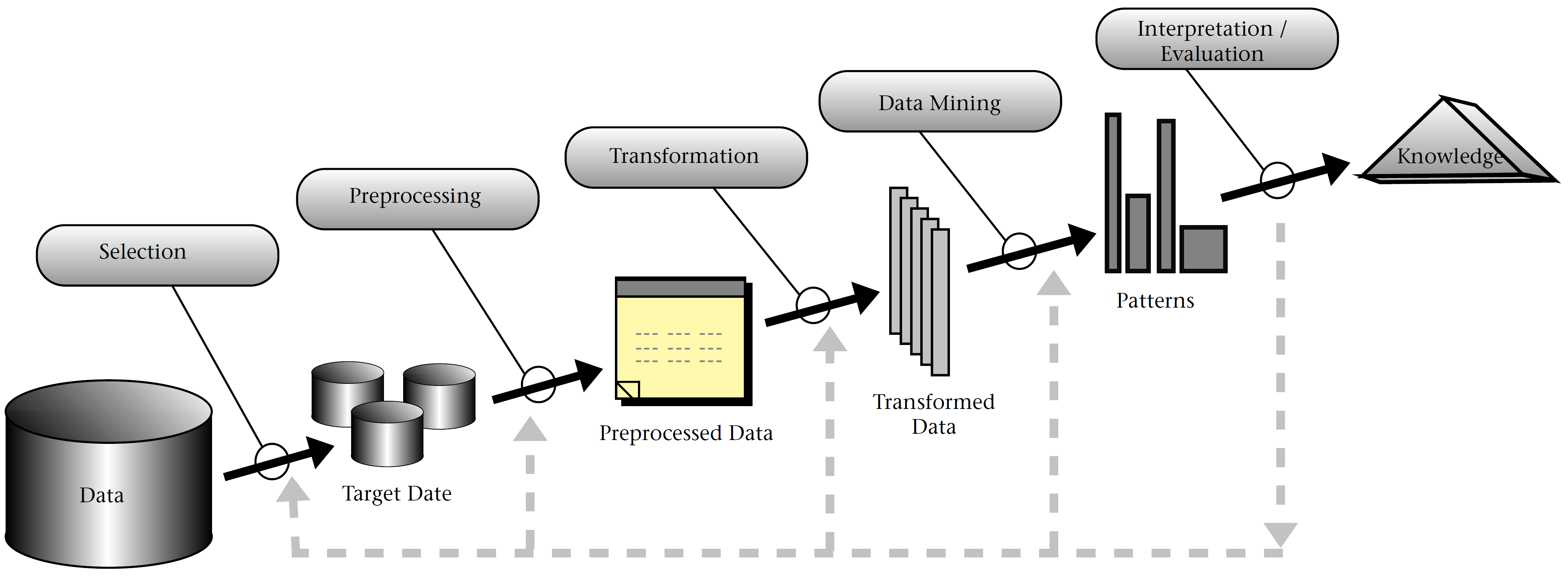}
	\label{fig:KDDprocess}
	\caption{An Overview of the steps of the KDD Process (from \cite{Fayyad1996KDD})}
\end{figure}

The different phases in the KDD process are outlined in Figure \ref{fig:KDDprocess} where we see a large amount of unnecessary loops between steps and a lack of business guidance.

%The 5 A’s \citep{martinez2003optimizacion} is a process model that proposes the tasks that should be performed to develop a DM project and was one of CRISP-DM’s forerunners. Therefore, they share the same philosophy: 5 A’s proposes the tasks but does not suggest how they should be performed. Its life cycle is similar to the one proposed in CRISP-DM. 

Several other process models and methodologies have been developed using the KDD approaches as a basis. The \emph{Human-Centered Approach to Data Mining} is presented in \citep{Brachman1996PKD,gertosio2004knowledge}. This proposal involves a holistic understanding of the entire Knowledge Discovery Process and involves eight steps: human resource identification, problem specification, problem specification, data prospecting, methodology identification, data preprocessing, pattern discovery, and knowledge post-processing. It considers people’s involvement and interpretation in each process and put emphasis on that the target user is the data engineer.

\emph{SEMMA} \citep{SEMMA}, which that stands for Sample, Explore, Modify, Model and Assess, is the methodology that the SAS institute\footnote{\url{http://www.sas.com}} proposed for developing DM products. Although it is a methodology, it is based only on the technical part of the project and integrated into SAS tools such as \emph{Enterprise Miner}. Unlike the former KDD process, SEMMA is not an open process and can only be used in these tools. The steps of SEMMA are mainly focussed on the modeling tasks of DM projects, leaving the business aspects. The steps are the following: sample, explore, modify, model and assess.

The two models by \citep{cabena1998discovering} and \citep{anand1998decision,anand1998data,buchner1999internet} are based on KDD process with not big differences and with similar features. The former structures the process in a different number of steps (business objectives determination, selection, preprocessing and transformation, data mining, analysis of results and assimilation of knowledge) and was used more in the marketing and sales domain, this being one of the first process models which took into account the business objectives. For its part, the latter process model is adapted to web mining projects and focused on an online customer (incorporating the available operational and materialized data as well as marketing knowledge). The model consists of eight steps: human resource identification, problem specification, problem specification, data prospecting, methodology identification, data preprocessing, pattern discovery, and knowledge post-processing. Although it provides a detailed analysis for the initial steps, it  does not include information on using the obtained knowledge.

The \emph{Two Crows} \cite{edelstein1998introduction} is a process model proposed by Two Crows Consulting\footnote{\url{http://twocrows.com/}} and takes advantage of some insights from (first versions of) CRISP-DM (before release). It proposes a non-linear list of steps (very close to the KDD phases), so it is necessary to go back and forth and . The basic steps of data mining for knowledge discovery are:  define business problem, build data mining database, explore data, prepare data for modeling, build model, evaluate model, deploy model and results.

%\begin{landscape}
%\end{landscape}

\subsection{Independent approaches}

There are some other independent approaches not related to the KDD original process. SPSS\footnote{\url{http://www.spss.com.hk/}} originally developed a data mining analysis cycle called the \emph{5 A’s Process} \citep{brunk1997mineset} included their data mining tool set. It involves five steps (Assess, Access, Analyse, Act and Automate) where the ``Automate'' step is the most relevant one and helps non-experts user to automate the whole process of DM applying already defined methods to new data. The main disadvantage is that the 5 A’s do not contain steps to understand the business objectives and to test data quality. The process was abandoned in 1999 when SPSS joined CRISP-DM consortium to develop the CRISP-DM process model.

In mid-1996, Motorola developed the $6-\sigma$ approach \citep{harry1998six} which emphasises measurement and statistical control techniques for quality and excellence in management. It is a well structured data-driven methodology for eliminating defects or quality control problems in manufacturing, service delivery, management, and other business activities, including data mining projects. That is done through the application of so-called ``Six Sigma DMAIC'' sequence of  steps (Define, Measure, Analyze,  Improve, and Control). This methodology has proven to be successful in companies such as IBM, Microsoft, General Electric, Texas Instrument or Ford.

\emph{KDD Roadmap} \citep{debuse2001building} is an iterative  data mining methodology methodology used in Witness Miner toolkit\footnote{\url{http://www.witnessminer.com/}} which uses a visual stream-based interface to represent routes through the KDD roadmap (consisting of eight steps: problem specification, resourcing, data cleansing, preprocessing, data mining, evaluation, interpretation and exploitation). The main contribution of KDD roadmap is the resourcing task which consist in the integration of databases from multiple sources to form the operational database.

\subsection{CRISP-DM: de facto standard}  

We focus on \emph{Cross Industry Standard Process for Data mining} (CRISP-DM) \citep{chapman2000crisp} as a process model because it is the ``de facto standard'' for developing DM and KD projects. In addition, CRISP-DM is the most used methodology for developing DM projects\footnote{CRISP-DM is still the top methodology for analytics, data mining, or data science projects according to \textit{kDnuggets}: \url{http://goo.gl/CYISan}}.  In general terms, CRISP-DM is a general purpose process model which is a freely available, industry independent, technology neutral, and it is said to be de facto standard for DM. 

\begin{figure}[h]
	\centering
		\includegraphics[width=0.60\textwidth]{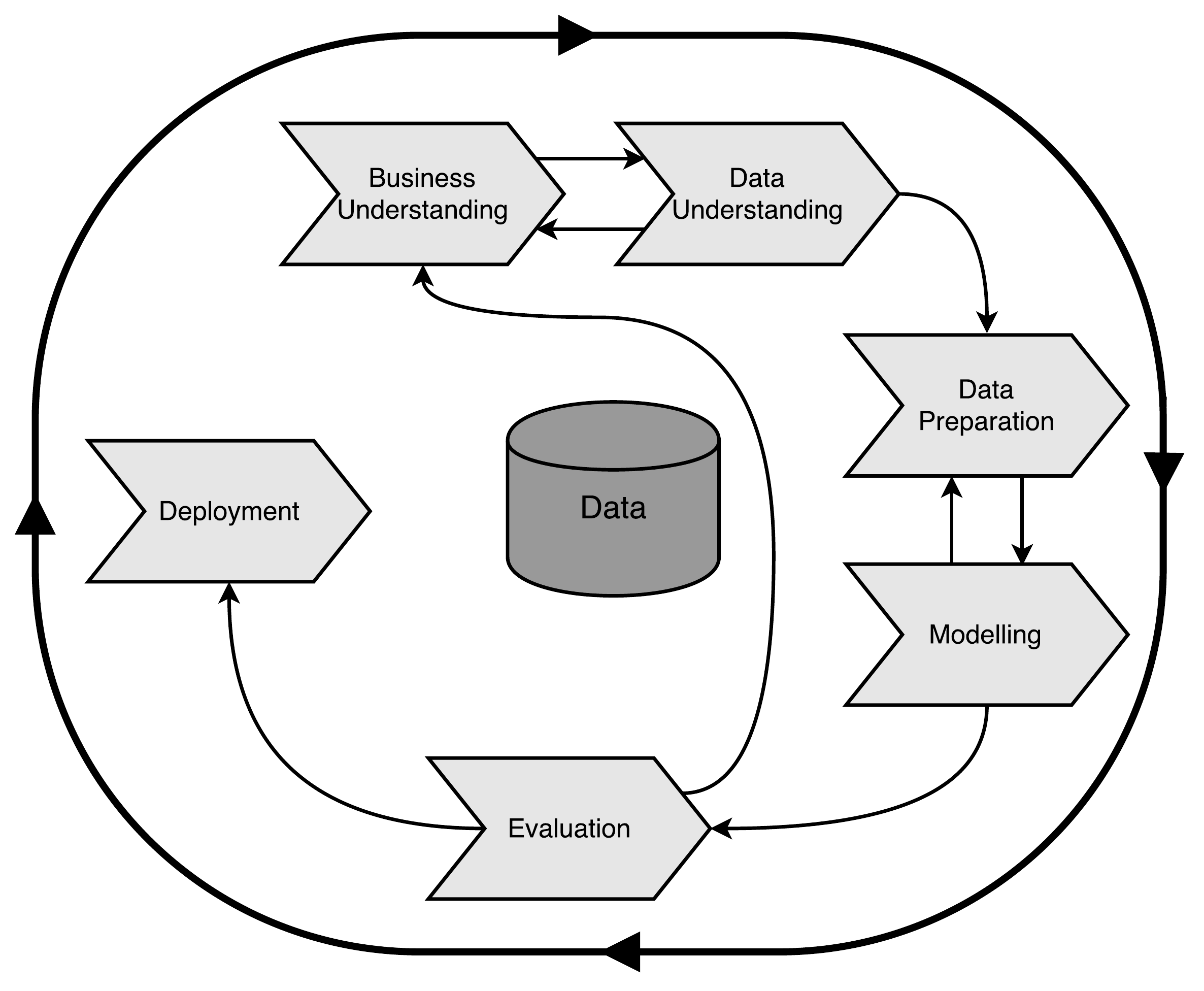}
	\caption{Process diagram showing the relationship between the different phases of CRISP-DM}
	\label{fig:CRISPDM-wikimedia}
\end{figure}

CRISP-DM, as a process model, provides an overview of the life cycle of a data mining project. It contains the  phases of a project, a set of tasks to be performed in each phase as well as the elements that are produced in each task (outputs) and the elements that are necessary to do a task (inputs). The life cycle of a data mining project consists of six phases (Figure \ref{fig:CRISPDM-wikimedia}) which sequence is not rigid: moving back and forth between different phases is always required and depends on the outcome of each phase which phase or which particular task of a phase, has to be performed next. The arrows indicate the most important and frequent dependencies between phases. The outer circle in Figure \ref{fig:CRISPDM-wikimedia} symbolizes the cyclical nature of data mining itself. Data mining is not over once a solution is deployed. Therefore data mining processes will benefit from the experiences of previous ones. 

In the following, we outline each phase briefly following the original reference model in \citep{chapman2000crisp}:

\begin{enumerate}
	\item \textbf{Business understanding}: This initial phase focuses on understanding the project objectives and requirements from a business perspective, then converting this knowledge into a data mining problem definition and a preliminary plan designed to achieve the objectives.
	
	\item \textbf{Data understanding}: The data understanding phase starts with an initial data collection and proceeds with activities in order to get familiar with the data, to identify data quality problems, to discover first insights into the data or to detect interesting subsets to form hypotheses for hidden information.
	
	\item \textbf{Data preparation}: The data preparation phase covers all activities to construct the final dataset from the initial raw data. Data preparation tasks are likely to be performed multiple times and not in any prescribed order. Tasks include table, record and attribute selection as well as transformation and cleaning of data for modeling tools.
	
	\item \textbf{Modeling}: In this phase, various modeling techniques are selected and applied and their parameters are calibrated to optimal values. Typically, there are several techniques for the same data mining problem type. Some techniques have specific requirements on the form of data. Therefore, stepping back to the data preparation phase is often necessary.
	
	\item \textbf{Evaluation}: At this stage the model (or models) obtained are more thoroughly evaluated and the steps executed to construct the model are reviewed to be certain it properly achieves the business objectives. A key objective is to determine if there is some important business issue that has not been sufficiently considered. At the end of this phase, a decision on the use of the data mining results should be reached
	
	\item \textbf{Deployment}: Creation of the model is generally not the end of the project. Even if the purpose of the model is to increase knowledge of the data, the knowledge gained will need to be organised and presented in a way that the customer can use it.

\end{enumerate}

Its final goal is to make the process repeatable, manageable and measurable (to be able to get metrics). CRISP-DM is usually referred as an informal methodology (although it does not provide the rigid framework, task/inputs/outputs specification and execution, evaluation metrics, or correctness criteria) because it provides the most complete tool set for DM practitioners. The current version includes the reference process model and implementation user guide defining phases, tasks, activities and deliverable outputs of these tasks. 

It is clear from Figure \ref{fig:EvolutionKDD} that CRISP-DM is the standard model and has borrowed principles and ideas from the most important models (KDD, SEMMA, Two Crowds,\dots) and has been the source for many later proposals. However, many changes have occurred in the business application of data mining since the former version of CRISP-DM was published: new data types and data mining techniques and approaches, more demanding requirements for scalability,  real-time deployment and large-scale databases, etc. The \emph{CRISP-DM 2.0} Special Interest Group (SIG) was established with the aim of meeting the changing needs of DM with and improved version of the CRISP-DM process. Normally this version should have appeared in 2007, but was finally discontinued.

However, other process models based on the original CRISP-DM approach have appeared. Cios et al.'s six-step 
discovery process  \citep{cios2000knowledge, cios2005trends} was first proposed in 2000 adapting the CRISP-DM model to the needs of the academic research community. The main extensions include, among others, improved (research-oriented) description of the steps, explicit feedback mechanisms, reuse of knowledge discovered between different domains, etc. The model consists of six steps: understanding the problem domain, understanding the data, preparation of the data, data mining, evaluation of the discovered knowledge and using the discovered knowledge. 

The \emph{RAMSYS} (RApid collaborative data Mining SYStem) \citep{moyle2001ramsys} is a methodology for developing DM and KD projects where several geographically diverse groups (nodes) work together on the same problem in a collaborative way. This methodology, although based on CRISP-DM  (same phases and generic tasks), emphasises collaborative work, knowledge sharing and communication between groups. Apart from the original CRISP-DM tasks, the RAMSYS methodology proposes a new task called model submission (modeling step), where the best models from each of the nodes are evaluated and delivered.

Finally, in 2015, IBM Corporation released \emph{ASUM-DM} (Analytics Solutions Unified Method for Data Mining/Predictive Analytics) a new methodology  which refines and extends CRISP-DM. ASUM-DM retained the “Analytical” activities and tasks of CRISP-DM but the method was augmented adding infrastructure, operations, deployment and project management sections as well as templates and guidelines. 

%Finally, the approaches presented by \citep{solarte2002proposed} (\emph{DMIE}) and by \citep{marban2009towar} are methodologies for data mining projects based on CRISP-DM, but oriented to industrial engineering domain. DMIE define five steps: analyse the organization, structure the work, develop the data model, implement the model, establish on-going support, where the latter step consists of a support and maintenance phase (data backups and maintenance,  model and software updates, \dots). For its part, the approach presented by \citep{marban2009towar} tries to include all the activities and tasks required (and not covered by CRISP-DM) in a (software) engineering standard process (based .
\section{Context-awareness and reuse of knowledge}\label{sec:context}

A major assumption in many machine learning and data mining algorithms is that the training and deployment data must be in the same contexts, namely, having the same feature space, distribution or misclassification cost. However, in many real-world applications, this assumption may not hold. Apart from having several different training contexts, there might also be many potential deployment contexts which differ from the training context(s) in one or more ways. An illustrative, nor exhaustive, list of context changes is shown is Table \ref{table:context-types}.

\begin{table}[h]
\centering
\resizebox{\textwidth}{!}{%
\begin{tabular}{ll}
\textbf{Context change}                                   & \textbf{Examples of parametrised context}           \\ \hline
Distribution shift (covariate, prior probability, concept) & Input or output variable distribution               \\
Costs and evaluation function                             & Cost proportion, cost matrix, loss function         \\
Data quality (uncertain, missing, or noisy information)   & Noise or uncertainty degree, missing attribute set  \\
Representation change, constraints, background knowledge  & Granularity level, complex aggregates, attribute set \\
Task change                                               & Binarised regression cut-off, bins                  
\end{tabular}%
}
\caption{Taxonomy of context change types and examples of their parametrisation.}
\label{table:context-types}
\end{table}

%First approach (universal model) [proactive]: Learn a model on extensive training data that covers all anticipated contexts.
%Second approach (transfer learning) [reactive]: Wait until the context changes and then learn a model in the  new context, transferring knowledge from the original context
%Third approach 3 (versatile model and reframing) [proactive + reactive]: Learn a model to solve a more general task than necessary for training context, reframe the model to deployment contexts

Many recent machine learning approaches have addressed the need to cope with context changes and reuse of learnt knowledge. Areas such as \emph{data shift} \cite{quinonero2009,moreno2012unifying,datashiftpatterns2014}, \emph{domain adaptation} \cite{jiang2008literature}, \emph{transfer learning} \cite{torrey2009transfer,pan2010survey}, \emph{transportability} \cite{bareinboim2012transportability}, \emph{meta-learning} \cite{giraud2004introduction}, \emph{multi-task learning} \cite{caruana1998multitask,thrun1996learning}, \emph{learning from noisy data} \cite{angluin1988learning,frenay2013classification}, \emph{context-aware computing} \cite{abowd1999towards}, \emph{mimetic models} \cite{blanco2006estimating}, \emph{theory revision} \cite{richards1991first}, \emph{lifelong learning} \cite{thrun2012learning} and \emph{incremental learning} \cite{Khreich:IS12}. Generally, in these areas the context change is analysed when it happens, rather than being anticipated, thus learning a model in the new context and reusing knowledge from the original context. 

\begin{figure}[!ht]
	\centering
		\includegraphics[width=0.70\textwidth]{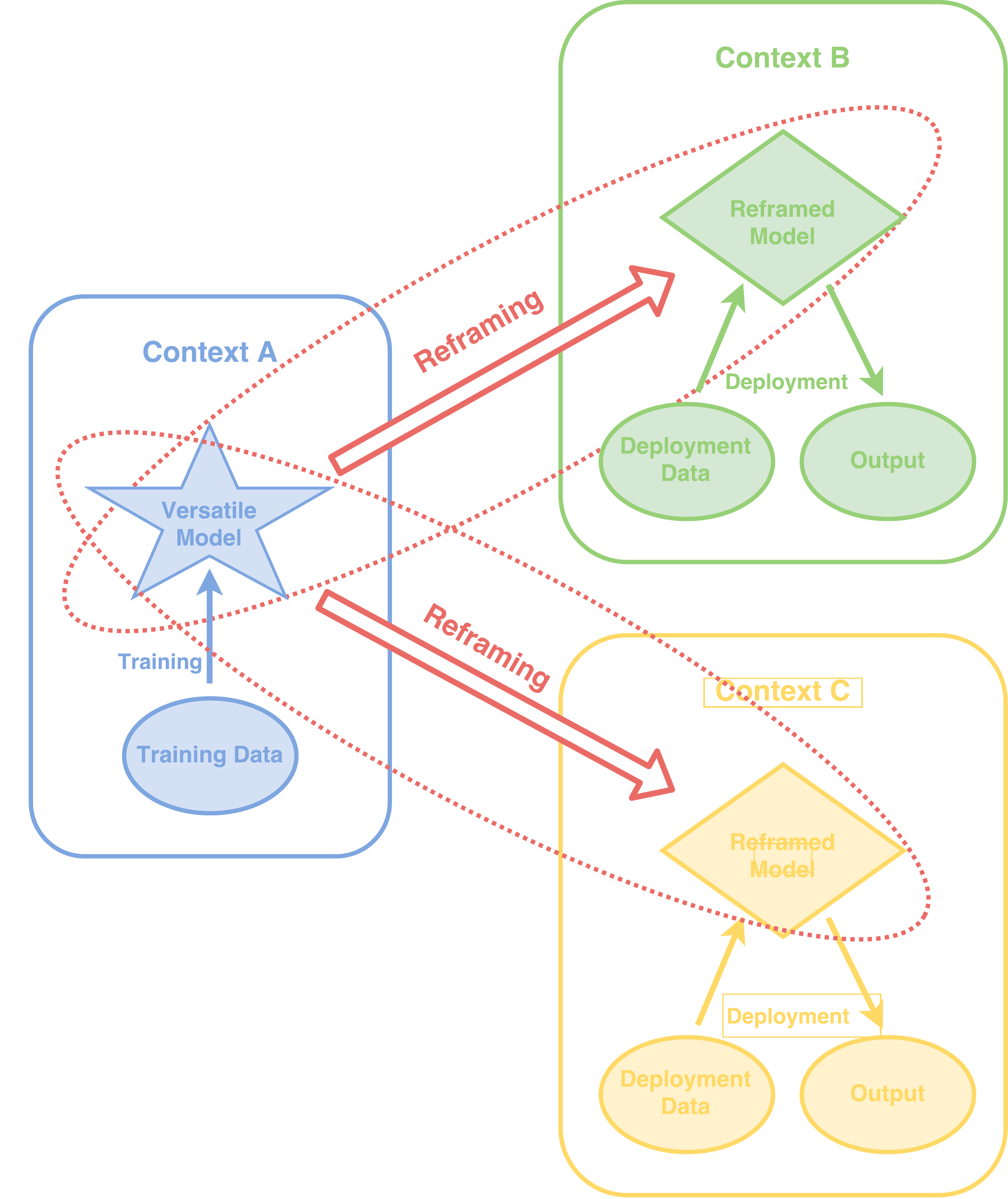}
	\caption{Operating contexts, models and reframing. The model on the left is intentionally more versatile than strictly necessary for context A, in order to ease its reframing to other contexts (e.g., B and C) without retraining it repeatedly.}
	\label{fig:Reframing}
\end{figure}

A more proactive way to deal with context changes is by constructing a {\em versatile} model, which has the distinct advantage that it is not fitted to a particular context or context change, and thus enables model reuse.  %Instead of training several specialised models for each particular operating context, it is more cost-effective to learn one general, versatile model, such as a scoring classifier outputting scores or probabilities, which can be adapted to several contexts through an appropriate procedure, such as the choice of a decision threshold. 
A new and generalised machine learning approach called \emph{Reframing} \cite{aicomHernandez16} addresses that. It formalises the expected context changes before any learning takes place, parametrises the space of contexts, analyses its distribution and creates versatile models that can systematically deal with that distribution of context changes. Therefore, the versatile model is reframed using the particular context information for each deployment situation, and not retrained or revised whenever the operating contexts change (see Figure \ref{fig:Reframing}). %This leead to a (more cost-effective) machine learning perspective in which models are not continuously retrained and re-assessed every time a change happens, but rather kept, enriched and validated in a long-term `model life-cycle'.
Rather than being an umbrella term for the above-mentioned related areas, reframing is a distinctive way of addressing context changes by anticipating them from the outset. \emph{Cost-sensitive learning} \cite{Elk01,turney2000types,chow1970optimum,tortorella2005roc,pietraszek2007use,vanderlooy2006analysis} and \emph{ROC analysis and cost plots} \cite{metz1978basic,flach2003decision,fawcett2006introduction,flach2010roc,drummond-and-Holte2006,ICML11CoherentAUC,ICML11Brier,hernandez2012unified,MLJ2013} can be seen as areas where reframing has been commonly used in the past, and generally restricted to binary classification.

%Generally speaking, there are three main alternatives for adapting to context: \emph{retraining}, \emph{revising} and \emph{reframing}. The former 

Generally speaking, the process of preparing a model to perform well over a range of different operation contexts involves a number of challenges:

\begin{itemize}
	\item \textbf{Reuse of learnt knowledge}: Models are required to be more general and adaptable to changes in the data distribution, data representation, associated costs, noise, reliability, background knowledge, etc. This naturally leads to a perspective in which models are not continuously retrained and re-assessed every time a change happens, but rather kept, enriched and validated in a long-term model life-cycle. This lead us to the concept of versatile models, able to generalise over a range of contexts.
	
	\item \textbf{Variety of contexts and context changes}: The process of preparing and devising a versatile model to perform well over a range of operating contexts (beyond the specific context in which the model was trained) involves to deal with a number of different possible context changes that are commonly observed in machine learning applications: distribution shift \cite{datashiftpatterns2014, moreno2012unifying,quinonero2009}, cost and evaluation function \cite{Elk01,turney2000types,chow1970optimum,pietraszek2007use,tortorella2005roc,vanderlooy2006analysis}, data quality \cite{frenay2013classification}, representation change \cite{multidimensional, AEPIA15}, constrains, background knowledge, task change \cite{scheirer2013toward,Orallo2016DMKD},\dots
	
	\item \textbf{Context-aware approaches for machine learning: Retraining vs. Revision  vs. Reframing trilemma}: \emph{Retraining} on the training data is very general, but there are many cases where it is not applicable. For instance, the training data may have been lost or may not exist (e.g., training models that have been created or modified by human experts) or may be prohibitively large (if deployment must work in restricted hardware), or the computational constraints do not allow retraining for each deployment context separately. Retraining on the deployment data can work well if there is an abundance of deployment data, but often the deployment data are limited, unsupervised or simply non-existent. 	A common alternative to retraining is \emph{revision}, \cite{DRa92,richards1991first} where parts of the model are patched or extended according to a new context (detection of novelty or inconsistency of the new data with respect to the existing model). It is especially natural as a result of an incremental learning \cite{Khreich:IS12} or lifelong learning \cite{thrun2012learning}. Finally, \emph{reframing}, as said above, is a context-aware approach that reuses the model trained in the training context by subjecting it to a reframing procedure that takes into account the particular deployment context . 	
	
	\item \textbf{Context-aware performance evaluation and visualisation}: When the context is constant, conventional context insensitive performance metrics can be used to evaluate how a model performs for that context. However, when we use the same model for several contexts we need context-aware performance metrics \cite{PRL09,ICML11Brier,MLJ2013,ICML11CoherentAUC,hand2009measuring,Hernandez-Orallo2015,JMLR2012,drummond-and-Holte2006,Lo:Wang:Wang:Lin:TransOnM11,RROC2013,Xu:Kusner:Weinberger:Chen:Chapelle:JMLR14,MissingAttributes2014,bij2003regression,fawcett2006introduction,flach2010roc,flach2003decision,metz1978basic}.
	
\end{itemize}

\noindent These challenges require a change of methodology. If we have to be more anticipative with context, we need a process model where context is present from the very beginning, and the analysis, identification and use of context (changes) must be part of several stages. This is what CASP-DM undertakes.

\begin{figure}[!ht]
	\centering
		\includegraphics[width=0.8\textwidth]{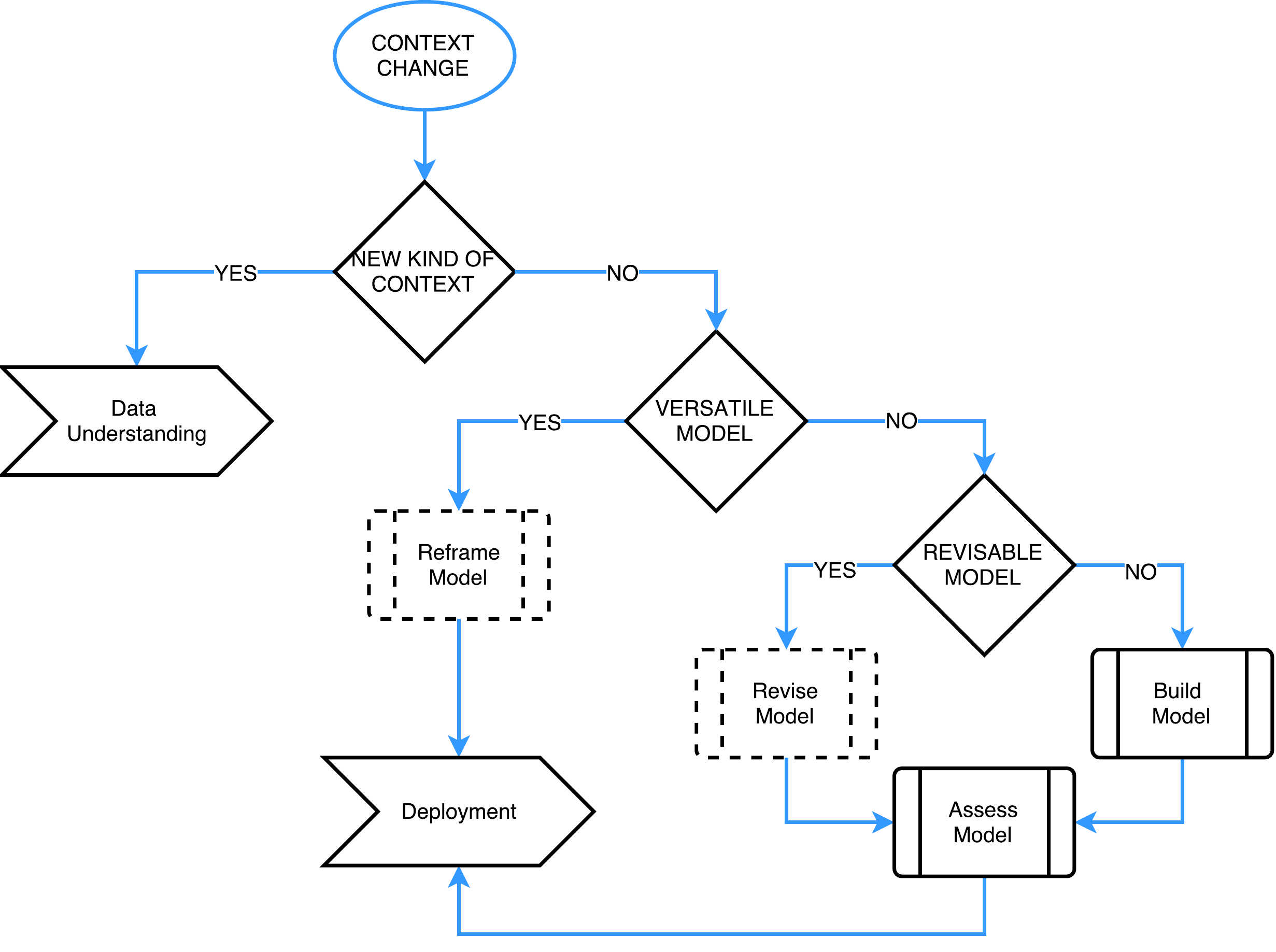}
	\caption{Tasks to be accomplished when there is a context change.}
	\label{fig:fcA}
\end{figure}

\section{CASP-DM}\label{sec:casposo}

\begin{figure}[!ht]
	\centering
		\includegraphics[width=0.70\textwidth]{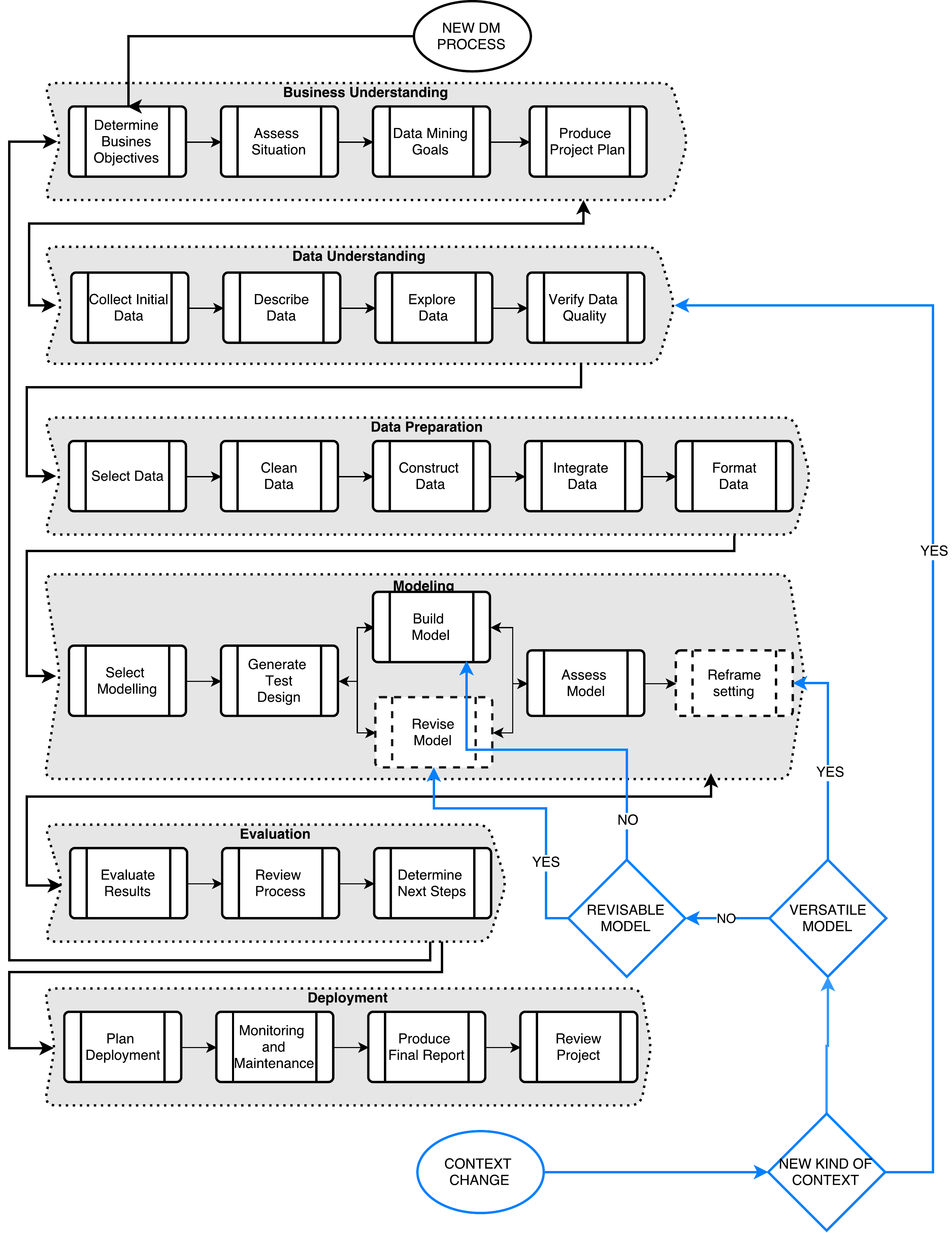}
	\caption{Complete view of the CASP-DM tasks to be completed whenever (1) a new context-aware DM project starts; or (2) a context change needs to be addressed.}
	\label{fig:fcB}
\end{figure}

CASP-DM, which stands for Context-Aware Standard Process for Data Mining, is the proposed extension of CRISP-DM for addressing specific challenges of machine learning and data mining for context and model reuse handling. CASP-DM model inherits flexibility and versatility from the CRISP-DM life cycle  and put more emphasis in that the sequence of phases is not rigid: context changes may affect different tasks so it should be possible to move to the appropriate phase. This is illustrated in Figures \ref{fig:fcA} (simplified) \ref{fig:fcB} (complete), where a flow chart shows which tasks in the CASP-DM process model should be completed whenever a context change needs to be addressed.

\begin{figure}[!ht]
	\centering
		\includegraphics[width=0.8\textwidth]{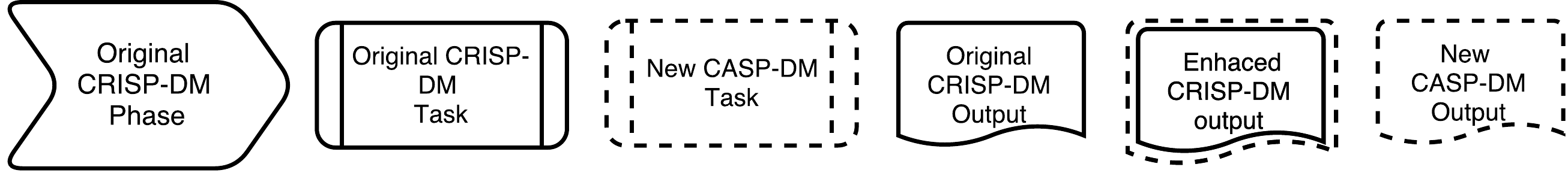}
	\caption{Legend of the different representation of original and new/enhanced tasks and outputs.}
	\label{fig:legend}
\end{figure}

In this section we  overview the life cycle of a DM project by putting emphasis on those new and enhanced tasks and outputs that have to do with context and model reuse handling (Figure \ref{fig:legend}). Enhanced or new tasks/outputs are shown in {\color{red!50!black}{\textbf{dark red}}}. Furthermore, a running example of model reuse with bike rental station data (MoreBikes) \cite{KullLU15} will be used to illustrate how CASP-DM is applied in a real environment.

\subsection{Business Understanding}

\begin{figure}[!ht]
	\centering
		\includegraphics[width=1.00\textwidth]{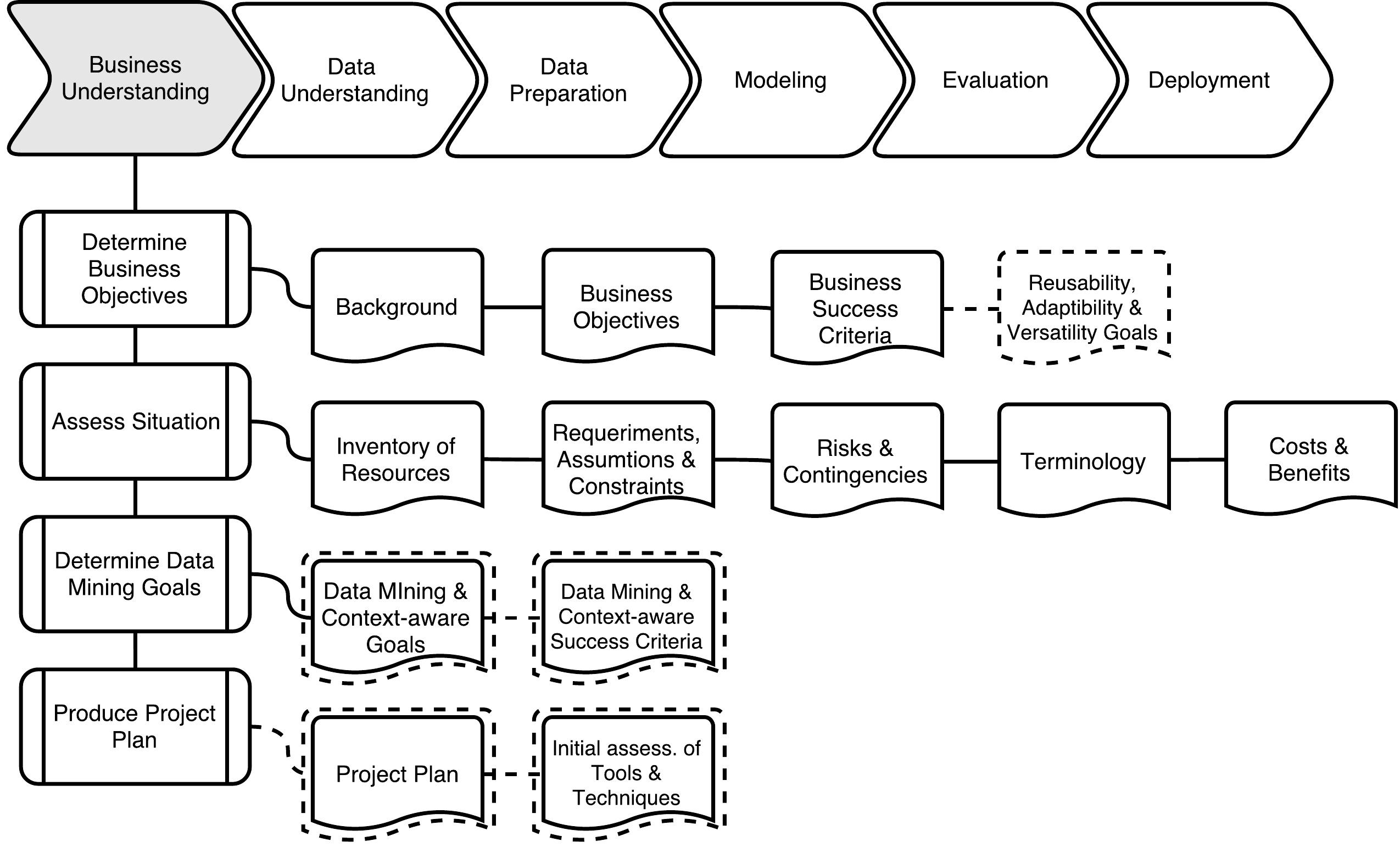}
	\caption{Phase 1. Business understanding: tasks and activities for context-awareness}
	\label{fig:BusinessUnderstanding}
\end{figure}

The CASP-DM first phase ``Business understanding'' (as well as the second phase ``Data understanding'') is the phase where the data mining project is being understood, defined and conceptualized. The rest  phases are implementation-related phases, which aim to resolve the tasks being set in the first phases. As in the original CRISP-DM, the implementation phases are highly incremental and iterative where the lessons learned during the process and from the deployed solutions can benefit subsequent data mining processes.

The initial phase focuses on understanding the project objectives and requirements from a business perspective, then converting this knowledge into a data mining problem definition and a preliminary plan designed to achieve the objectives. Adapting this phase to address context changes and model reuse handling involves: (1) adding new specialized tasks for identifying long term reusability business goals (whether the business goals involve reusability, adaptability, and versatility) w.r.t. context changes, (2) determining both data mining goals and success criteria when we address a context-aware data mining problem (which type of context-aware technique should be used depends on what aspects of the model are reusable in other contexts) and, finally, (3) perform an initial assessment of available context-aware techniques and update the project plan describing the intended plan for achieving the data mining goals and thereby achieving the reusability, adaptability, and versatility business goals. The plan should specify the steps to be performed during the rest of the project, including the initial identification of contexts (changes), and the reframing techniques (I/O, structural) to deal with them. 

\subsubsection{Determine business objectives}
\begin{itemize}
	\item \textbf{Task}: The first task is to thoroughly understand, from a business perspective, what the client really wants to accomplish, and thus try to gain as much insight as possible into the business goals for data mining. For that it is necessary to gather background information about the current business situation, document specific business objectives and agree upon criteria used to determine data mining success from a business perspective. 
			
		\item \textbf{Outputs}:
		\begin{itemize}
				\item \textbf{Background}: Record the information that is known about the organization’s business situation: determine organizational structure, identify the problem area and describe any solutions currently used to address the business problem
				\item \textbf{Business objectives}: Describe the customer’s primary objective agreed upon by the project sponsors and other business units affected by the results
				\item \textbf{Business success criteria}: Define the nature of business success for the data mining project from the business point of view. This might be as precisely as possible and able to be measured objectively.
				\item {\color{red!50!black}{\textbf{Reusability, Adaptability and Versatility Goals}}}: Identify, from a business long-term perspective, which are the prerequisites and future perspectives: whether the business goals involve reusability, adaptability, and versatility (i.e., should our solution procedure perform well over a range of different operating contexts?). 
				
		\end{itemize}
\end{itemize}

\begin{texample}
          
					\textbf{Finding Business Objectives} \par
					Adaptive reuse of learnt knowledge is of critical importance in the majority of knowledge-intensive  application  areas,  particularly  when  the  context  in  which the learnt model operates can be expected to vary from training to deployment. The  MoReBikeS challenge (Model Reuse with Bike Rental Station Data) organised as the ECML-PKDD 2015 Discovery Challenge \#1 \cite{kullmorebikes}, is focused on model reuse and context change.
					
					The  MoReBikeS  challenge  was  carried  out  in  the  framework  of  historical
bicycle  rental  data  obtained  from  Valencia,  Spain.  Bicycles  are  continuously taken from and returned to rental stations across the city. Due to the patterns in demand some stations can become empty or full, such that more bikes cannot be  rented  or  returned.  To  reduce  the  frequency  of  this  happening,  the  rental company has to move bikes from full or nearly full stations to empty or nearly empty stations. Therefore the task is to predict the number of available bikes in every bike rental stations 3 hours in advance. There are at least two use cases for such predictions. 
					
					\begin{itemize}
						\item First, a specific user plans to rent (or return) a bike in 3 hours time and wants to choose a bike station which is not empty (or full).
						\item Second, the company wants to avoid situations where a station is empty or full and therefore needs to move bikes between stations. For this purpose they need to know which stations are more likely to be empty or full soon. 
					\end{itemize}
						
					\textbf{Context-awareness}:  Information from older stations should be used to improve performance on younger ones. In future, new stations will be planned every few months, but probably the growth is getting faster.

\end{texample}

\subsubsection{Assess situation}
\begin{itemize}
	\item \textbf{Task}: Once the goal is clearly defined, this task involves more detailed fact-finding about all of the resources, constraints, assumptions and other factors that should be considered in determining the data analysis goal and project plan.
		
		\item \textbf{Outputs}:
		\begin{itemize}
				\item \textbf{Inventory of resources}: Accurate list of the resources available to the project, including: personnel, data sources, computing resources and software. 
				\item \textbf{Requirements, assumptions and constraints}: List all requirements of the project (schedule of completion, security and legal restrictions, quality, etc.), list the assumptions made by the project (economic factors, data quality assumptions, non-checkable assumptions about the business upon which the project rests, etc.) and list the constraints on the project (availability of resources, technological and logical constraints, etc.). 								
				\item \textbf{Risks and contingencies}:  List of the risks or events that might occur to delay the project or cause it to fail (scheduling, financial, data, results, etc.) and list of the corresponding contingency plans.
				\item \textbf{Terminology}: Compile a glossary of technical terms (business and data mining terminology) and buzzwords that need clarification. 
				\item \textbf{Costs and benefits}: Construct a cost-benefit analysis for the project (comparing the estimated costs with the potential benefit to the business if it is successful).
		\end{itemize}
\end{itemize}

\begin{texample}
         		\textbf{Assessing the Situation} \par
					
					One of the first tasks the consultant faces is to assess the company’s resources for data mining. 
					\begin{itemize}
						\item 	\textbf{Data}. Since this is an established company, there is plenty of historical information from stations as well as information about the current status, time of the day/week/year,  geographical data, weather conditions, etc.  
					\end{itemize}
\end{texample}

\subsubsection{Determine data mining goals}
\begin{itemize}
	\item \textbf{Task}: Translate business goals (in business terminology) into data mining goal reality (in technical terms). 
		
		\item \textbf{Outputs}:
		\begin{itemize}
				\item {\color{red!50!black}{\textbf{Data mining and context-aware goals}}}: Describe the type of data mining problem. Initial exploration of how the different contexts are going to be used. Describe technical goals. Describe the desired outputs of the project that enables the achievement of the business objectives. 
				\item {\color{red!50!black}{\textbf{Data mining and context-aware success criteria}}}: Define the criteria for a successful outcome to the project in technical terms: describe the methods for model and context assessment, benchmarks, subjective measurements, etc.
				
		\end{itemize}
\end{itemize}

\begin{texample}

         	\textbf{Data Mining Goals} \par
					Bike rental company needs to move bikes around to avoid empty and full stations. This can be done more efficiently if the numbers of bikes in the stations are predicted some hours in advance. The quality of such predictions relies  heavily  on  the  recorded  usage  over  long  periods  of  time.  Therefore,  the prediction quality on newly opened stations is necessarily lower. The goals for the study are:
					
					\begin{itemize}
						\item Use historical information about bike availability in the stations. In this challenge we explore a setting where there are 200 stations which have been running for more than 2 years and 75 stations which have just been open for a month. 
						\item Reuse the models learned on 200 ``old'' stations in order to improve prediction performance on the 75 ``new'' stations. Combine information from similar stations to build improved models. Hence, this challenge evaluates prediction performance on the 75 stations.
						\item By predicting the number of bikes in the new stations (3 hours in advance), the bike rental company will be able to move bikes around to avoid empty and full stations.
					\end{itemize}
\end{texample}

\subsubsection{Produce project plan}
\begin{itemize}
	\item \textbf{Task}: Describe the intended plan for achieving the data mining goals and thereby achieving the business goals. The plan should specify he project of the business goals, data mining goals (reusability, adaptability, and versatility), resources, risks, and schedule for all phases of data mining as well as include an initial selection of tools and techniques.
		
		\item \textbf{Outputs}:
		\begin{itemize}
				\item {\color{red!50!black}{\textbf{Project plan}}}: List the stages to be executed in the project, together with duration, resources required, inputs, outputs and dependencies. Where possible make explicit the large-scale iterations in the data mining process, for example repetitions of the modeling and evaluation phases.
								
				\item {\color{red!50!black}{\textbf{Initial assessment of tools and techniques}}}:  At the end of the first phase, the project also performs an initial assessment of tools and techniques, including the initial identification of contexts (changes) and the context-aware techniques to deal with them.	
				
		\end{itemize}
\end{itemize}

\begin{texample}
    			\textbf{MoReBikeS  Example---Assessing Tools and Techniques} \par
					After setting the project plan for the study, an initial selection of tools and techniques should be made taking into account contexts and context changes:
					\begin{itemize}
						\item In this challenge, context is the combination of station and time. It should be advisable to use model combination, and retraining on sets of similar station. 
					\end{itemize}
\end{texample}

%%%%%%%%%%%%%%%%%%%%%%%%%%%%%%%%%%%%%%%%%%%%%%%%%%%%%%%%%%%%%%%%%%%%%%%%%%%%%%%%%%%%%%%%%%%%%%%%%%%%%%%%%%%%%%%%%%%%%%%%%%%%%%%%%%%%%%%%%%%%
%%%%%%%%%%%%%%%%%%%%%%%%%%%%%%%%%%%%%%%%%%%%%%%%%%%%%%%%%%%%%%%%%%%%%%%%%%%%%%%%%%%%%%%%%%%%%%%%%%%%%%%%%%%%%%%%%%%%%%%%%%%%%%%%%%%%%%%%%%%%
%%%%%%%%%%%%%%%%%%%%%%%%%%%%%%%%%%%%%%%%%%%%%%%%%%%%%%%%%%%%%%%%%%%%%%%%%%%%%%%%%%%%%%%%%%%%%%%%%%%%%%%%%%%%%%%%%%%%%%%%%%%%%%%%%%%%%%%%%%%%
%%%%%%%%%%%%%%%%%%%%%%%%%%%%%%%%%%%%%%%%%%%%%%%%%%%%%%%%%%%%%%%%%%%%%%%%%%%%%%%%%%%%%%%%%%%%%%%%%%%%%%%%%%%%%%%%%%%%%%%%%%%%%%%%%%%%%%%%%%%%%%%%%%%%%%%%%%%%%%%%%%%%%%%%%%%%%%%%%%%%%%%%%%%%%%%%%%%%%%%%%%%%%%%%%%%%%%%%%%%%%%%%%%%%%%%%%%%%%%%%%%%%%%%%%%%%%%%%%%%%%%%%%%%%%%%%%%%%%%%%
%%%%%%%%%%%%%%%%%%%%%%%%%%%%%%%%%%%%%%%%%%%%%%%%%%%%%%%%%%%%%%%%%%%%%%%%%%%%%%%%%%%%%%%%%%%%%%%%%%%%%%%%%%%%%%%%%%%%%%%%%%%%%%%%%%%%%%%%%%%%
%%%%%%%%%%%%%%%%%%%%%%%%%%%%%%%%%%%%%%%%%%%%%%%%%%%%%%%%%%%%%%%%%%%%%%%%%%%%%%%%%%%%%%%%%%%%%%%%%%%%%%%%%%%%%%%%%%%%%%%%%%%%%%%%%%%%%%%%%%%%
%%%%%%%%%%%%%%%%%%%%%%%%%%%%%%%%%%%%%%%%%%%%%%%%%%%%%%%%%%%%%%%%%%%%%%%%%%%%%%%%%%%%%%%%%%%%%%%%%%%%%%%%%%%%%%%%%%%%%%%%%%%%%%%%%%%%%%%%%%%%

\subsection{Data Understanding}

\begin{figure}[!ht]
	\centering
		\includegraphics[width=1.00\textwidth]{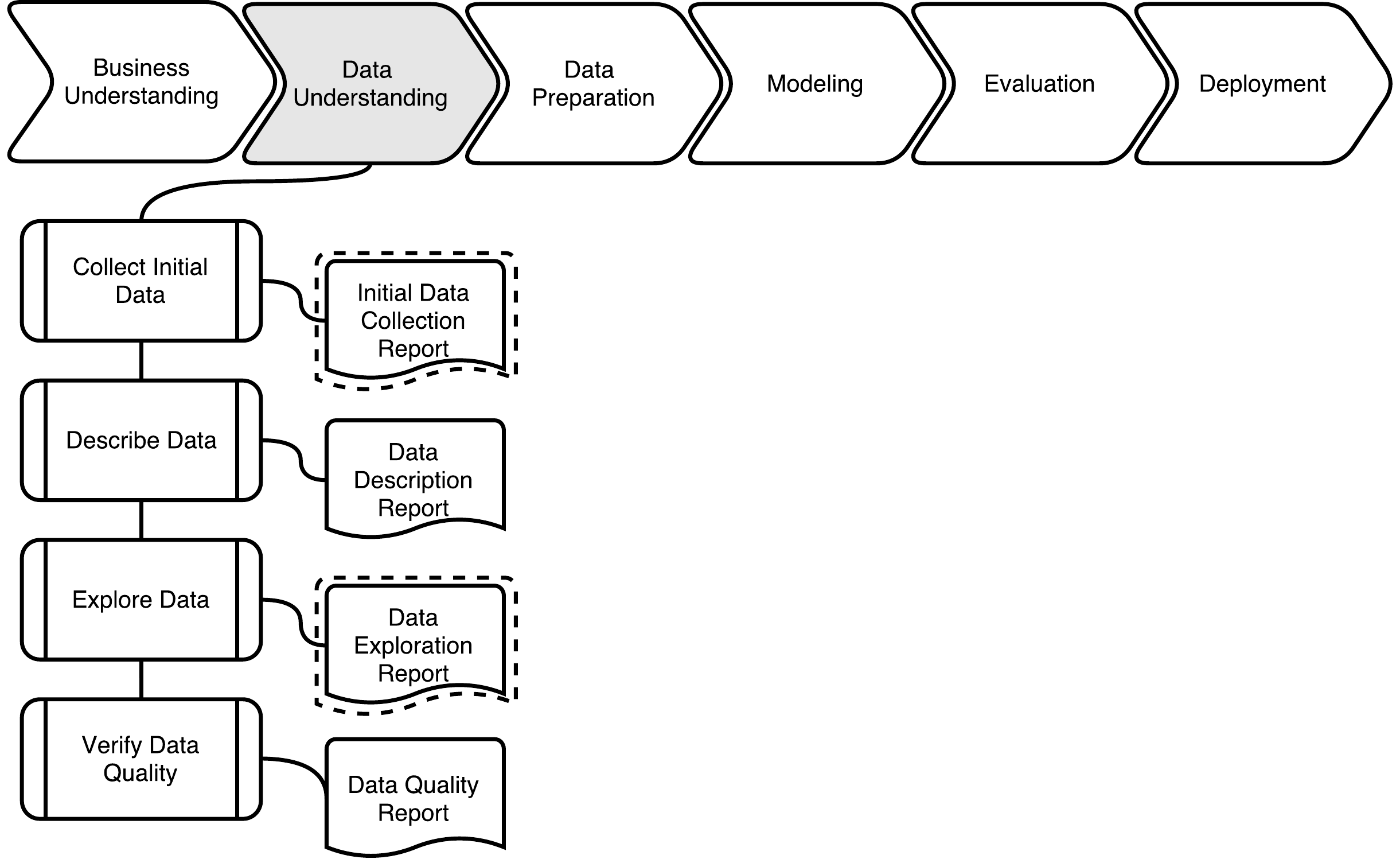}
	\caption{Phase 2. Data Understanding: tasks and activities for context-awareness}
	\label{fig:DataUnderstanding}
\end{figure}

The CRISP-DM phase 2 ``Data understanding'' involves an initial data collection and proceeds with activities that enable you to become familiar with the data, identify data quality problems, discover first insights into the data, and/or detect interesting subsets to form hypotheses regarding hidden information.

To adapt this second phase to address the new needs, we have to enhance the initial data collection task in order to be able represent different relevant contexts. Through a further data exploration we should be also able to contribute to or refine the data description, quality reports and information about context representation, and feed into the transformation and other data preparation steps needed for further analysis. %Finally, we should explore possible data quality changes between training and deployment contexts. 

\subsubsection{Collect initial data}
\begin{itemize}
	\item \textbf{Task}: Acquire the data (or access to the data) listed in the project resources. This initial collection includes data  integration if acquired from multiple data sources. Describe attributes (promising, irrelevant, \dots), quantity and quality of data. Collect sufficiently rich raw data to represent possibly different relevant contexts. Collect sufficiently rich raw data to represent possibly different relevant contexts.

	\item \textbf{Outputs}: 
	\begin{itemize}
				\item {\color{red!50!black}{\textbf{Initial Data Collection Report}}}: Describe data collected: describe attributes (promising, irrelevant, \dots), quantity and quality of data and identify relevant contexts.			
		\end{itemize}
	\end{itemize}
	
\begin{texample}
          \textbf{Initial Data Collection} \par
				A procedure to store the number of bikes in all stations every hour has been set up. The  gathered data provides information about 275 bike rental stations in Valencia over a period of 2.5 years (from 01/06/2012 to 31/01/2015). For each hour in this period the data specified the median number of available bikes during that hour in each of the stations. The dataset was complemented with weather information about the same hour (temperature, relative humidity, air pressure, amount of precipitation, wind directions, maximum and mean wind speed).
				
				The bike rental data for Valencia have been obtained from \url{http://biciv.com}, weather information from the Valencian Regional Ministry for the Environment (\url{http://www.citma.gva.es/}) and holiday information from \url{http://jollyday.sourceforge.net/}.
\end{texample}

\subsubsection{Describe data}

\begin{itemize}
	\item \textbf{Task}: Describe the properties of the acquired data and report on the results. This includes the amount of data (consider sampling), value types, records, fields, coding schemes, etc.

	\item \textbf{Outputs}: 
	\begin{itemize}
				\item {\textbf{Initial Data Collection Report}}: Write description report in order to share the findings about the data. 
		\end{itemize}
	\end{itemize}

\begin{texample}
          \textbf{Describing Data} \par
				
					There are 24 given features in total which can be divided to 4 categories:
					
					\begin{itemize}
						\item \textbf{Facts of stations}. The facts of stations provided in the data set include the station ID, the latitude, the longitude and the number of docks in that station. All these properties for one station do not change over time.
						\item \textbf{Temporal information}. The timestamp of a data entry consists of eight
 fields:  ``Timestamp''  in  terms  of  seconds  from  the  UNIX  epoch,  ``Year'', ``Month'', ``Day'', ``Hour'', ``Weekday'', ``Weekhour'', and ``IsHoliday'' which indicates whether the day is a public holiday. These features are giving overlapping temporal information, we only need a subset of them to represent a time point. The ``Timestamp'' is actually including information of ``Year'', ``Month'', ``Day'', ``Hour'',``Weekday'' and ``Weekhour'', whereas ``Weekday'' and ``Hour'' also can be deduced by ``Weekhour''. Only ``IsHoliday'' is independent to any of others.
						\item \textbf{Weather}. This set of features include ``windMaxSpeed'', ``windMeanSpeed'', ``windDirection'', ``temperature'', ``relHumidity'', ``airPressure'', ``Precipitation''.  One  major  observation  of  weathers  is  that  all  the  values  of  all  the seven fields share among all stations.
						\item \textbf{Counts and their statistics}.  This  set  of  features  relates  to  the  target value  directly.  First  of  all,  ``bikes 3h ago''  gives  the  target  value  of  the  3-hour-earlier time point at a station. The full profile features use all previous data points of the same ``Weekhour'' to obtain long term statistics for each ``Weekhour'' in each station, accordingly the short profile features only use at most four previous data points to obtain short-term statistics. The long-term statistics of the 200 old stations only have very small changes over time in contrast to the short-term ones
						
					\end{itemize}
					 The target variable is ``bikes'' and it is a non-negative integer representing the median number of available bikes during the respective hour in the respective rental station.
					%The data contain 4 station features, 8 time features, 7 weather features, 1 task-specific feature and 4 profile features plus 1 target variable. The target variable is 'bikes' and it is a non-negative integer representing the median number of available bikes during the respective hour in the respective rental station. There are 4 features about the station (station, latitude, longitude, numDocks), 8 features about timepoint (timestamp, year, month, day, hour, weekday, weekhour, isHoliday) and 7 features about weather (windMaxSpeed.m.s, windMeanSpeed.m.s, windDirection.grades, temperature.C, relHumidity.HR, airPressure.mb, precipitation.l.m2) which do no differ across stations.
\end{texample}

\subsubsection{Explore data}
\begin{itemize}
	\item \textbf{Task}: This task addresses data mining and context-aware goals through querying, visualization, and reporting techniques over the data and how they may contribute/refine the initial (business or DM) goals, data transformation/preparation.... Among others, this analysis include distribution of key attributes, looking for errors in the data, relationships between pairs or small numbers of attributes, results of simple aggregations, properties of significant sub-populations, and simple statistical analyses. 
	
	\item \textbf{Outputs}: 
	\begin{itemize}
				\item {\color{red!50!black}{\textbf{Data exploration report}}}:  Describe results  of this task including (possibly using graphs and plots) first findings, initial hypothesis, explorations about contexts, particular subsets of relevant data and attributes and their impact on the remainder of the project. 
		\end{itemize}
	\end{itemize}
	
\begin{texample}
          \textbf{Exploring Data} \par
					
					A lot of work should be done in this stage in the bike scenario. Taking pieces of domain knowledge and checking whether they hold and identify interesting patterns. 
					We can see that different stations clearly exhibit different daily patterns. Most obviously, there are stations that tend to be full in the night and emptier during the day. Essentially these are stations that are on the outer areas of the city, and the bikes are used during the day to travel into more central parts of the city. There are also stations that exhibit the opposite pattern. These stations are left empty at night, since the operators know that the will fill up during the day as people travel into the city. There are of course stations that fall between these two extremes.
\end{texample}

\subsubsection{Verify data quality}
\begin{itemize}
	\item \textbf{Task}: Examine the quality of the data: coding or data errors, missing values, bad metadata, measurement errors and other types of inconsistencies that make analysis difficult.
	
	\item \textbf{Outputs}: 
	\begin{itemize}
				\item \textbf{Initial Data Collection Report}: List and describe the results of the data quality verification (is correct?, contain errors?, missing values?,  how common are they?) and list possible solutions. 
		\end{itemize}
	\end{itemize}

\begin{texample}
          \textbf{Verifying Data Quality} \par
					
					Some of the issues encountered include missing values in the profile information about the station that could be ignored. Timepoint features also have missing values and only the timepoints with existing values are used.
\end{texample}

\subsection{Data Preparation}

\begin{figure}[!ht]
	\centering
		\includegraphics[width=1.00\textwidth]{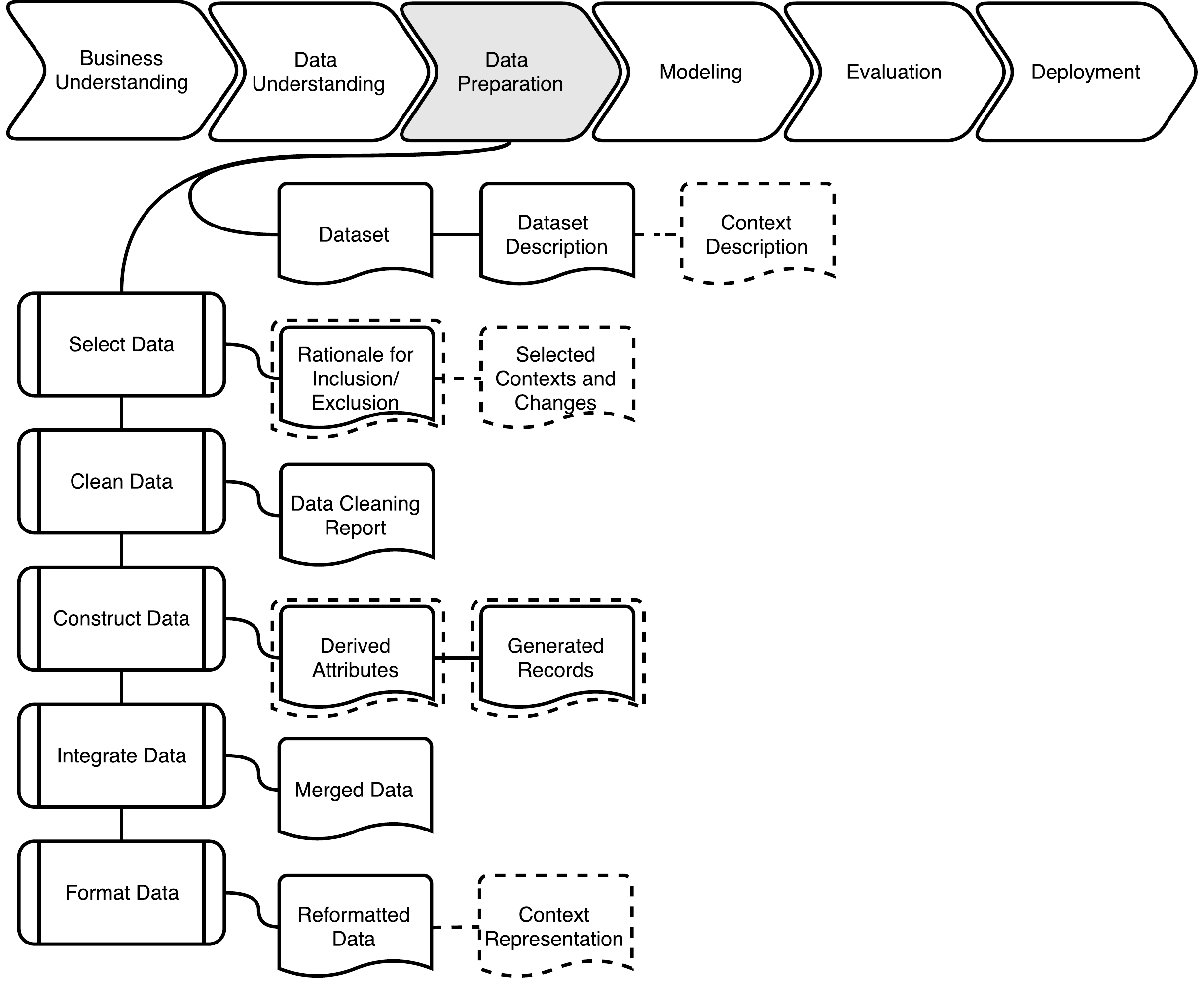}
	\caption{Phase 3. Data Preparation: tasks and activities for context-awareness}
	\label{fig:DataPreparation}
\end{figure}

The CRISP-DM phase 3 ``Data preparation'' covers all activities needed to construct the final dataset (data that will be fed into the modelling tool(s)) from the initial raw data. Data preparation tasks are likely to be performed multiple times and not in any prescribed order. In fact, it is estimated that data preparation usually takes 50-70\% of a project’s time and effort. Tasks include table, record, and attribute selection, as well as transformation and cleaning of data for modelling tools. The original CRISP-DM task ``select data'' had limitations for practical application in complex domains (e.g. multidimensional) since it is mostly assumed for single-table static data format. Furthermore, it lacks activities to handle data wrangling, data conversion, data sampling, data unification, etc. 

Select data has been enhanced with feature extraction, resolution change and dimensionality reduction techniques to define possible attribute sets for modelling activities. Furthermore a selection of contexts and context changes relevant to the data mining goals should be done by selecting data which cover the selected contexts and changes.  Enhanced constructive data preparation operations has been added to derive context-specific and context-independent attributes.  The integration of data from multiple tables or records to create new records or values should be also updated with data from different contexts. Finally, data formatting for specific data mining techniques need to include the context representation.

\subsubsection{Select data}
\begin{itemize}
	\item \textbf{Task}: Based upon the initial data collection conducted in the previous CRISP-DM phase, you are ready to decide on the data to be used for analysis. Note that data selection covers selection of records (rows) as well as attributes (columns) in a table.
	
	\item \textbf{Outputs}: 
	\begin{itemize}
				\item {\color{red!50!black}{\textbf{Rationale for inclusion/exclusion}}}:  List the data and context to be included/excluded and the reasons for these decisions. 
				\item {\color{red!50!black}{\textbf{Selected  contexts and changes}}}:  Select contexts and context changes relevant to the data mining goals, ignore the others. Select data to cover the selected contexts and changes. 
		\end{itemize}
	\end{itemize}
	
\begin{texample}
          \textbf{Selecting Data} \par
					Many of the decisions about which data and attributes to select have already been made in earlier phases of the data mining process.  Contexts are modelled as parameters (station and timestamp) and both need to be modelled later (using all available data).
			
					%\textbf{Features to remove}:
					%\begin{itemize}
						 %\item ``Year'' and ``Month'' are fixed to October 2014 in the training data set, there is nothing to learn from them. 
						%\item Accordingly ``Timestamp'' is too general to distinguish different temporal information and has too many possible values to clearly indicate similarity between different time points. 
						%\item The actual meanings of full profile features are different from 200 old stations to 75 new features, because the historical data used to calculate the long term statistics are spanning over 2 years for the 200 old stations but only several weeks for the new 75 stations.
						%
						%\textbf{Features to keep}:						
						%\begin{itemize}
							%\item ``IsHoliday'' doesn't overlap with any other temporal features and gives extra information in addition to the timestamp. 
							%\item Since ``Timestamp'' is already removed, ``Day'' has become a
%temporal feature without any alternative and probably includes some periodical information.
							%\item ``Bikes of 3 hours ago'' %and short profile features, 
							%unlike long term statistics, these features of 75 new stations are aligned with the 200 old stations, so they can be very informative.
							%\item The existing linear models only keep ``temperature'' among seven weather features, which implies it is a useful information.
						%\end{itemize}
						%
					%\end{itemize}
\end{texample}

\subsubsection{Clean Data}
\begin{itemize}
	\item \textbf{Task}: Clean and solve problems in the data chosen to include for the analysis. This tasks aims at raising the data quality to the level required by the selected analysis techniques.
	
	\item \textbf{Outputs}: 
	\begin{itemize}
				\item \textbf{Data Cleaning Report}: Report data-cleaning efforts (missing data, data errors, coding inconsistencies, missing data and bad metadata) for tracking alterations to the data and in order for future data mining projects to be benefited. 
			
		\end{itemize}
	\end{itemize}

\begin{texample}
          \textbf{Selecting Data} \par
					The bike rental company uses the data cleaning process to address the problems noted in the data quality report.						
				\begin{itemize}
						\item \textbf{Missing data}. The missing values are ignored in all profile calculations, i.e. only the timepoints with existing values are averaged.
				\end{itemize}
\end{texample}

\subsubsection{Construct data}
\begin{itemize}
	\item \textbf{Task}: This task includes constructive data preparation operations such as the production of derived attributes or entire new records, or transformed values for existing attributes.
	
	\item \textbf{Outputs}: 
	\begin{itemize}
				\item {\color{red!50!black}{\textbf{Derived attributes}}}: Derived attributes are new attributes that are constructed from one or more existing attributes in the same record. Derive context-specific and context-independent attributes. 
				\item {\color{red!50!black}{\textbf{Derived attributes}}}: Describe the creation of completely new records. Generate new data to force context-invariance (e.g., rotated images in deep learning).
			
		\end{itemize}
	\end{itemize}

\begin{texample}
            \textbf{Constructing Data} \par
Several new parameters are created to be added to the profiles of each station:
						
						\begin{itemize}
							\item There is one feature about the number of bikes in the station 3 hours ago: ``bikes 3h ago''. The profile variables are calculated from earlier available timepoints on the same station.
							\item The ``full profile bikes'' feature is the arithmetic average of the target variable ``bikes'' during all past timepoints with the same weekhour, in the same station.
							
							\item The ``full profile 3h diffbikes'' feature is the arithmetic average of the calculated feature ``bikes-bikes 3h ago'' during all past timepoints with the same weekhour, in the same station.
							
							\item The ``short *'' profile is the same as the full profiles except that it only uses past 4 timepoints with the same weekhour. If there are less than 4 such timepoints then all are used.  
						\end{itemize}
\end{texample}

\subsubsection{Integrate data}
\begin{itemize}
	\item \textbf{Task}: These are methods whereby information is combined from multiple sources. There are two basic methods of integrating data: merging two data sets with similar records but different attributes or  appending two or more data sets with similar attributes but different records. 
	\item \textbf{Outputs}: 
	\begin{itemize}
				\item {\color{red!50!black}{\textbf{Merged data}}}: This includes: merging tables together into a new table; aggregation of data (summarising information) from multiple records and/or tables and integrating data from relevant contexts
							
		\end{itemize}
	\end{itemize}

\begin{texample}
          \textbf{Selecting Data} \par
					With multiple data sources (bike station' historical data, bike stations' current status, weather conditions and profile data) it is necessary to integrate all data.

\end{texample}

\subsubsection{Format data}
\begin{itemize}
	\item \textbf{Task}: This task involves checking whether certain techniques require a particular format or order to the data. Therefore syntactic modifications have to be made to the data (without changing its meaning).
	
	\item \textbf{Outputs}: 
	\begin{itemize}
				\item \textbf{Reformatted data}: Syntactic changes made to satisfy the requirements of the specific modeling tool. Examples: change the order of the attributes and/or records, add identifier, remove commas from within text fields, trimming values, etc.
				\item {\color{red!50!black}{\textbf{Context representation}}}: Select context representation. (How are the contexts going to be represented in the data (parametrisation; as-feature vs as-dataset)?) 
							
		\end{itemize}
	\end{itemize}

\begin{texample}
          \textbf{Context representation} \par
					As commented in previous phases, context in this challenge is represented by means of parameters, concretely station identifier and timestamp.
\end{texample}

%%%%%%%%%%%%%%%%%%%%%%%%%%%%%%%%%%%%%%%%%%%%%%%%%%%%%%%%%%%%%%%%%%%%%%%%%%%%%%%%%%%%%%%%%%%%%%%%%%%%%%%%%%%%%%%%%%%%%%%%%%%%%%%%%%%%%%%%%%%%%%%%%%%%%%%%%%%%%%%%%%%%%%%%%%%%%%%%%%%%%%%%%%%%%%%%%%%%%%%%%%%%%%%%%%%%%%%%%%%%%%%%%%%%%%%%%%%%%%%%%%%%%%%%%%%%%%%%%%%%%%%%%%%%%%%%%%%%%%%%%%%%%%%%%%%%%%%%%%%%%%%%%%%%%%%%%%%%%%%%%%%%%%%%%%%%%%%%%%%%%%%%%%%%%%%%%%%%%%%%%%%%%%%%%%%%%%%%%%%%%%%%%%%%%%%%%%%%%%%%%%%%%%%%%%%%%%%%%%%%%%%%%%%%%%%%%%%%%%%%%%%%%%%%%%%%%%%%%%%%%%%%%%%%%%%%%%%%%%%%%%%%%%%%%%%%%%%%%%%%%%%%%%%%%%%%%%%%%%%%%%%%%%%%%%%%%%%%%%%%%%%%%%%%%%%%%%%%%%%%%%%%%%%%%%

\subsection{Modelling}

\begin{figure}[!ht]
	\centering
		\includegraphics[width=1.00\textwidth]{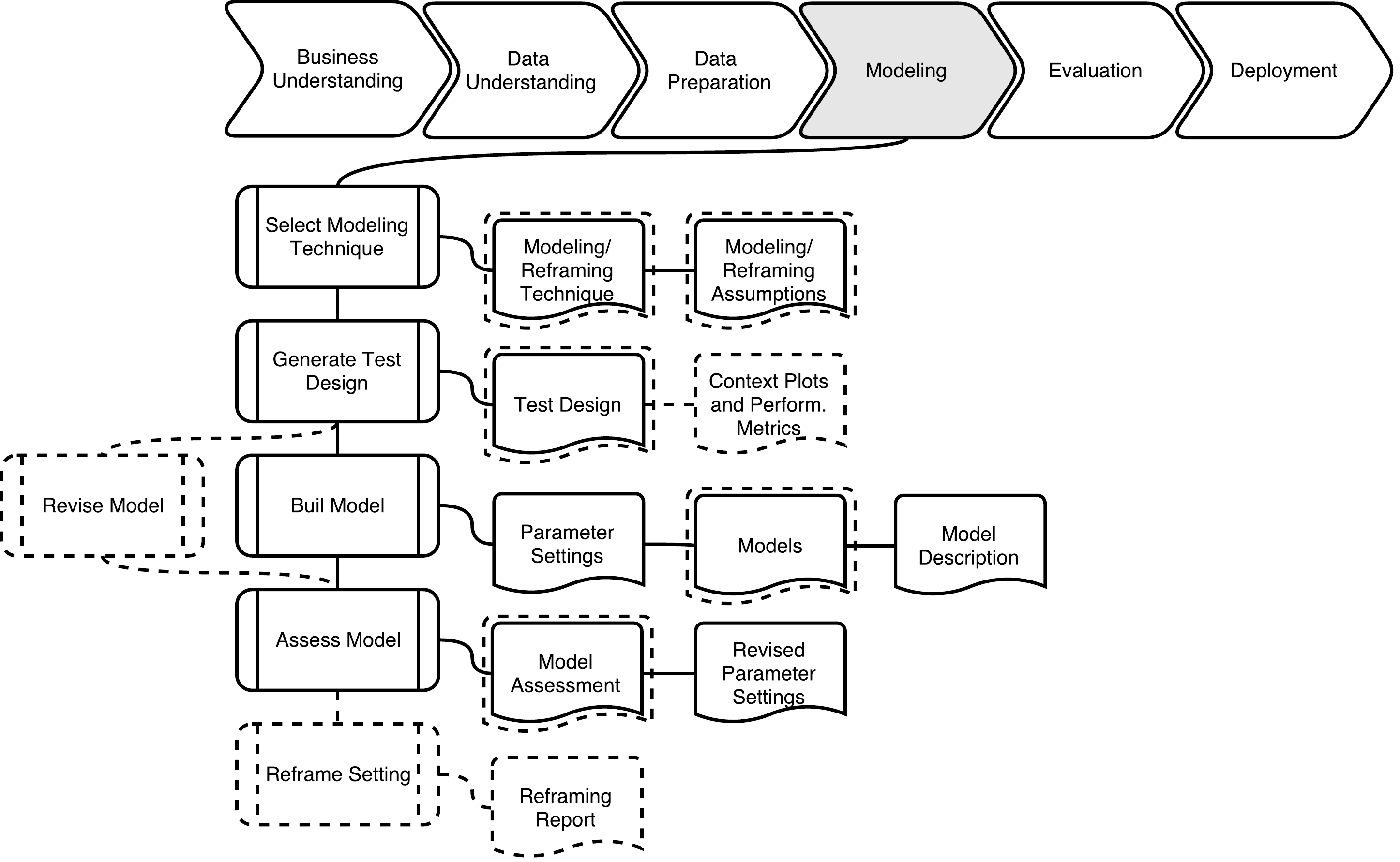}
	\caption{Phase 4. Modelling: tasks and activities for context-awareness}
	\label{fig:Modeling}
\end{figure}

In this phase, various modeling techniques are selected and applied, and their parameters are calibrated to optimal values. Typically, there are several techniques for the same data mining problem type and this phase is usually conducted in multiple iterations. Some techniques have specific requirements on the form of data, so going back to the data preparation phase is often necessary. %Furthermore, there is an iteration (not originally defined) within this phase between the tasks “Build Model” and “Assess Model”.

A new optional branch of reframe-based subtasks and deliverables has been added for selecting the modelling technique. Therefore, we clearly differentiate between classical modelling techniques and reframing techniques. Furthermore, enhanced procedures for testing the versatile model’s quality and validity (context plots and performance metrics) has been added. Specific reframing tools are needed to build the versatile model. A new general task “REVISE MODEL” for handling model revision in incremental or lifelong learning data mining tasks. Furthermore, a new general task “REFRAME SETTING” has been added in this phase in order to decide which type of reframing should be used (over the versatile model) depending on what aspects of the model are reusable in other contexts. This task will be performed to adapt a versatile model w.r.t. a context whenever the context changes. Finally, context-aware performance metrics are also needed to assess the versatile model.

\subsubsection{Select modelling technique}
\begin{itemize}
	\item \textbf{Task}: As the first step in modelling, select the actual modeling technique that is to be used. Although it has been selected a tool during the ``Business Understanding'' phase, this task refers to the specific modeling technique, e.g., decision-tree building with 5.0, or neural network generation with back propagation. Determining the most appropriate model will typically be based on the data types, data mining goals (scores, patterns, clusters, versatile model. etc.)
	\item \textbf{Outputs}: 
	\begin{itemize}
				\item {\color{red!50!black}{\textbf{Modeling technique}}}: Document the actual modeling technique that is to be used. In case context matters, Select the model and reframing couple, e.g.. scorer and score-driven or linear regression and continuous output reframing.
				\item {\color{red!50!black}{\textbf{Modeling assumptions}}}:  Many modeling techniques make specific assumptions on the data, e.g., all attributes have uniform distributions, no missing values allowed, class attribute must be symbolic etc. Record any such assumptions made.
		\end{itemize}
	\end{itemize}

\begin{texample}
          \textbf{Modeling Techniques} \par
				Because  of  that  lack  of  training data,  it  was  supposed  that historical  models about two-year  data  on  the  old  bike  stations  would  yield  better  predictions  that  the  scarce  training  data.  Each station is thus characterised by 6 linear models to predict the number of bikes, corresponding to 6 different subsets of the features ``ikes 3h ago'', ``full profile bikes'', ``full profile 3h diffikes'', ``short profile bikes'', ``short profile 3h diff bikes'' and temperature. All the subsets included ``bikes 3h ago'' but differed based on which profile features they used (3 options: full profiles, short profiles, or all profiles), and whether they used temperature (2 options: yes or no). We use regression modelling techniques techniques in order to handle data with outliers. We also impute Missing Values by median/mode. %The obtained 200 $\times$ 6 models were provided to the participants.

							%\begin{itemize}
		%\item \textbf{short}: ``bikes 3h ago'', ``short profile 3h diff bikes'', ``short profile bikes''.
					%\item \textbf{short temp}: ``bikes 3h ago'', ``short profile 3h diff bikes'', ``short profile bikes'', ``temperature.C''.
					%\item \textbf{full}: ``bikes 3h ago'', ``full profile 3h diff bikes'', ``full profile bikes''.
					%\item \textbf{full temp}: ``bikes 3h ago'', ``full profile 3h diff bikes'', ``full profile bikes'', ``temperature.C''
					%\item \textbf{short full}: ``bikes 3h ago'', ``short profile 3h diff bikes'', ``short profile bikes'', ``full profile 3h diff bikes'', ``full profile bikes''
					%\item  \textbf{short full temp}: ``bikes 3h ago'', ``short profile 3h diff bikes'', ``short profile bikes'', ``full profile 3h diff bikes'', ``full profile bikes'', ``temperature.C''.
				%\end{itemize}

				Therefore ,  the  hypothesis  made  was  that  the  closest  old  stations  to  the  target stations were most capable to predict future use of  those new stations given the  different  models  for  the  other  200  stations. For  that  reason,  distance  seems to be   a  crucial point in weighting the predictions of the given models. 
\end{texample}

\subsubsection{Generate test design }
\begin{itemize}
	\item \textbf{Task}: Before we actually build a model, we should consider how the model’s results will be tested. Therefore we need to generate a procedure or mechanism to test the model’s quality and validity (describing the criteria for goodness of a model (i.e., error rate) and defining the data on which these criteria will be tested.

	\item \textbf{Outputs}: 
	\begin{itemize}
				\item {\color{red!50!black}{\textbf{Test design}}}: Describe the intended plan (i.e., how to divide the available dataset) for training, testing and evaluating the models. 
				\item {\color{red!50!black}{\textbf{Context plot and performance metrics}}}: Decide how the context changes can be evaluated (e.g, by using artificial data). Identify proper metrics to evaluate reframing efficiency.
		\end{itemize}
	\end{itemize}

\begin{texample}
          \textbf{Test Design} \par
						
						As already commented, the stations are first splitted randomly into 200 training stations and 75 test stations. The time period was splitted into training period (01/06/2012 to 31/10/2014) and three-months test period (01/11/2014 to 31/01/2015). The last month of the training period (01/10/2014 to 31/10/2014) we referred to as the deployment period. We trained 6 different linear models (more details in the previous task) for each of the 200 training stations on the training period. The participants are provided with the trained models, with the data from the one-month deployment period for all 200+75 stations, and with the data from the training period from 10 training stations out of 200.

			The criteria by which the models are assessed is the Mean Absolute Error (MAE) in three-month test period across 50 test stations, with different forecasting windows, grouped by length of history and perhaps some meta-information about the station. 
\end{texample}

\subsubsection{Build Model}
\begin{itemize}
	\item \textbf{Task}: Run the modelling tool on the prepared dataset to create one or more models.

	\item \textbf{Outputs}: 
	\begin{itemize}
				\item \textbf{Parameter settings}: Most modeling techniques have a a large number of parameters that can be adjusted. List the parameters and their chosen value, along with the rationale for the choice of parameter settings.
				\item {\color{red!50!black}{\textbf{Models}}}: These are the actual models produced by the modeling tool, not a report.
				\item {\color{red!50!black}{\textbf{Model description}}}: Describe the resultant model. Report on the results of a model and any meaningful conclusion, document any difficulties or inconsistencies encountered with their meanings.
		\end{itemize}
	\end{itemize}

\begin{texample}
					\textbf{Model Building} \par
					
%For each of the 200 training stations we learned 6 linear models to predict the number of bikes, corresponding to 6 different subsets of the features (seen above) ``bikes\_3h\_ago'', ``full\_profile\_bikes'', ``full\_profile\_3h diff\_bikes'', ``short\_profile\_bikes'',``short\_profile\_3h\_diff\_bikes'' and temperature. All the subsets included ``bikes\_3h\_ago'' but differed based on which profile features they used (3 options: full profiles, short profiles, or all profiles), and whether they used temperature (2 options: yes or no). 

In choosing the models to be reused there is a good range of approaches: perform an analysis to select for each test station one model out of the given 1200, and used that model for prediction. Select multiple models and averaged over these. Use a weighted average of model. Finally, and following the retraining approach, it should be decided not to use the given model and trained new models.

\begin{itemize}
	\item \textbf{Reframe} version: a possible solution to the problem consists of combining the predictions of the K nearest stations among the old stations (1:200) to the target stations (201:275) using the weighted arithmetic mean.  On one hand, these predictions are calculated applying the best model---in terms of MAE---for each old station (1:200). On the other hand, the K nearest neighbours were obtained by comparing each target stations (201:275) to all the old stations (1:200) in terms of the Euclidean distance between them. Then, the K closest old stations to one target station were selected as its K nearest neighbours. In doing so for every target station (201:275), their K nearest neighbours were discovered among the old stations (1:200). The Euclidean distance between the target station and its neighbours is used to weight the influence of their predictions on the final prediction. Finally, this summation was divided by the sum of the k Euclidean distances from each neighbour (among the K nearest neighbours) to the target station on the test data. In doing so, the final prediction value was obtained from k predictions taken into account in a different importance according to their proximity to the target station.
	\item \textbf{Retraining} version: it consists on using the data of the roughly 2.5 year long period between 2012 and 2014 for 10 docking stations in the city of Valencia as well as the one month partial training data  provided for 190 other stations throughout the city.
\end{itemize}
\end{texample}

\subsubsection{Revise Model}
\begin{itemize}

	\item \textbf{Task}: Once we have built a model and as a result of an incremental learning or lifelong learning, the model needs to be revised(patched or extended) because of some novelty or inconsistency of the new data is detected with respect to the existing model. This can be extended to context changes, provided we can determine when the context has changed significantly to deserve a revision process.

\end{itemize}

\subsubsection{Assess Model}
\begin{itemize}
	\item \textbf{Task}: 	For each model under consideration, we have to interpret them and make a methodical assessment according to the data mining success criteria, and the desired test design. Judge and discuss the the success of the application of modelling and discovery techniques technically. Rank the models used. 

	\item \textbf{Outputs}: 
	\begin{itemize}
				\item {\color{red!50!black}{\textbf{Model assessment}}}: Summarize results of this task by using evaluation chars, analysis nodes, cross-validation charts, etc.; list qualities of generated models (e.g., in terms of accuracy) and rank their quality in relation to each other. In context-aware tasks, compare with different scenarios, in particular retraining.
				\item {\textbf{Revised parameter settings}}: According to the model assessment, revise parameter settings and tune them for the next run in the Build Model task. Iterate model building and assessment until you strongly believe that you found the best model(s). Document all such revisions and assessments.

		\end{itemize}
\end{itemize}

\begin{texample}
    			\textbf{MoReBikeS Example---Model Assessment} \par
					The reframing solution, which selects multiple models and averages over these, is better than retraining.

\end{texample}

\subsubsection{Reframe setting}
\begin{itemize}
	\item \textbf{Task}: 	Which type of reframing technique should be used depending on what aspects of the model are reusable in other contexts? Taking into account the particular deployment context (if known), we distinguish three different kinds of reframing (which can be combined): \emph{output}, \emph{input} and \emph{structural} reframing. Thus, where a conventional, non-versatile model captures only such information as is necessary to deal with test instances from the same context, a versatile model captures additional information that, in combination with reframing, allows it to deal with test instances from a larger range of contexts.

	\item \textbf{Outputs}: 
	\begin{itemize}
				\item {\color{red!50!black}{\textbf{Kind of reframing}}}: Describe the kind of reframing (output, input or structural) to be applied over the versatile model.
		\end{itemize}
	\end{itemize}

\begin{texample}
    			\textbf{Reframe setting} \par
				
						 In choosing the models to be reused it has to be decided the criteria for model suitability for a given test station. These included: performance of the model in the test station during the deployment period; distance between the test station and the station of the model's origin; similarity between the time-series of the stations during the deployment period; and several combinations of these. 
\end{texample}

%%%%%%%%%%%%%%%%%%%%%%%%%%%%%%%%%%%%%%%%%%%%%%%%%%%%%%%%%%%%%%%%%%%%%%%%%%%%%%%%%%%%%%%%%%%%%%%%%%%%%%%%%%%%%%%%%%%%%%%%%%%%%%%%%%%%%%%%%%%%%%%%%%%%%%%%%%%%%%%%%%%%%%%%%%%%%%%%%%%%%%%%%%%%%%%%%%%%%%%%%%%%%%%%%%%%%%%%%%%%%%%%%%%%%%%%%%%%%%%%%%%%%%%%%%%%%%%%%%%%%%%%%%%%%%%%%%%%%%%%%%%%%%%%%%%%%%%%%%%%%%%%%%%%%%%%%%%%%%%%%%%%%%%%%%%%%%%%%%%%%%%%%%%%%%%%%%%%%%%%%%%%%%%%%%%%%%%%%%%%%%%%%%%%%%%%%%%%%%%%%%%%%%%%%%%%%%%%%%%%%%%%%%%%%%%%%%%%%%%%%%%%%%%%%%%%%%%%%%%%%%%%%%%%%%%%%%%%%%%%%%%%%%%%%%%%%%%%%%%%%%%%%%%%%%%%%%%%%%%%%%%%%%%%%%%%%%%%%%%%%%%%%%%%%%%%%%%%%%%%%%%%%%%%%%

\subsection{Evaluation}

\begin{figure}[!ht]
	\centering
		\includegraphics[width=1.00\textwidth]{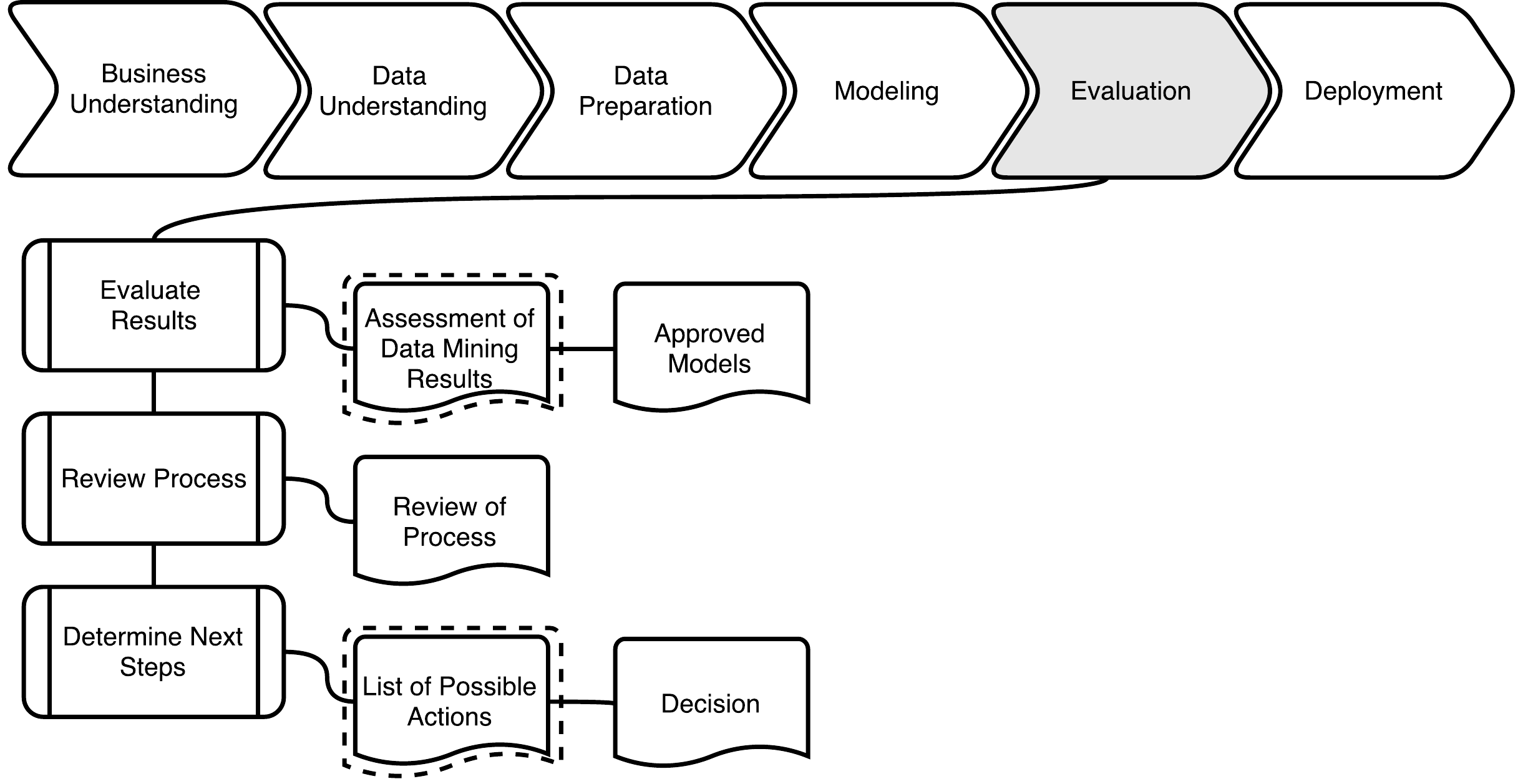}
	\caption{Phase 5. Evaluation: tasks and activities for context-awareness}
	\label{fig:Evaluation}
\end{figure}

Once you have built a model (or models) that, according to the evaluation task in the previous phase, appears to have high quality from a data analysis perspective, it is important to thoroughly evaluate it (perhaps going through the previous phases) to be certain the model properly achieves the business objectives. Therefore, this step requires a clear understanding of the stated business goals. A key objective is to determine how well results answer your organization’s business goals and whether there is some important business issue that has not been sufficiently considered.  At the end of this phase, a decision on the use of the data mining results should be reached.

Regarding context-awareness, in this phase we need an enhanced task for assessing whether the versatile model meets the business objectives in all the relevant contexts where they are to be deployed. Furthermore, we need to decide whether the versatile model is able to be reused and adapted to the deployment data or not.

%
%Deliverables for this task include two items:
%
    %Assessment of results (for business goals): Summarize the results with respect to the business success criteria that you established in the business-understanding phase. Explicitly state whether you have reached the business goals defined at the start of the project.
%
    %Approved models: These include any models that meet the business success criteria.

\subsubsection{Evaluate results}
\begin{itemize}
	\item \textbf{Task}: 	Unlike the previous evaluation steps which dealt with factors such as the accuracy and generality of the model, in this step we need to assess the degree to which the model meets the business objectives and, thus, this step requires a clear understanding of the stated business goals. We need to determine if there is some business reason why this model is deficient, if results are stated clearly, if there are  novel or unique findings that should be highlighted, if results raised additional questions, etc.

	\item \textbf{Outputs}: 
	\begin{itemize}
				\item {\color{red!50!black}{\textbf{Assessment of data mining results with respect to business success criteria}}}: Summarize assessment results in terms of business success criteria, interpret the data mining results, check the impact of result for initial application goal in the project, see if the discovered information is novel and useful, rank the results, state conclusions, check whether results cover all contexts relevant for the business success criteria, etc.
			
				\item {\textbf{Approved models}}: Select those (versatile) models which,  after the  previous assessment with respect to business success criteria, meet the the selected criteria.  
		\end{itemize}
	\end{itemize}
	
\begin{texample}
    			\textbf{Evaluating Results} \par
					Given a new rental station (deployment context), it is conceivable  that there might be some rental stations that are more similar to this station in terms of the daily usage patterns. Following this idea, the proposed here methods find the closest stations in terms of distance. The overall results indicate that the deployed methods are quite simple and easy to apply and can achieve a good performance when used to know which stations are more likely to be empty or full soon. 
					
					\begin{itemize}
					\item \textbf{New Questions}. The most important questions to come out of the study are:  How often the stations remain empty or full because of bad predictions? How much time is wasted in carrying bikes around because of bad predictions? Can we use different evaluation measures in modelling to achieve better results? How often do we need to retrain models?
					\end{itemize}
\end{texample}

\subsubsection{Review Process}
\begin{itemize}
	\item \textbf{Task}: Extra time for reflection on the successes and weaknesses of the process just completed. Although the resulting models appear to be satisfactory and to satisfy business needs,  it would be appropriate to do a more thorough review of the whole data mining process seeking for overlooked tasks and quality assurance issues. We should summarise activities and decisions made in each phase learning thus from your experience so that future data mining projects will be more effective.

	\item \textbf{Outputs}: 
		\begin{itemize}
				\item {\textbf{Review of process}}: Summarize the process review and all the activities and decisions for each phase. Give hints for activities that have been missed and/or should be repeated.
 		\end{itemize}
	\end{itemize}
				
\begin{texample}
					\textbf{MoReBikeS Example---Review Report} \par
					As a result of reviewing the process of the initial data mining project, the bike rental company has developed a greater appreciation of the interrelations between steps an its inherent ``backtracking'' nature. Furthermore, the company has learn that model reuse between similar stations is appropriate when historical data is not provided or does not exists.
\end{texample}

\subsubsection{Determine next steps}
\begin{itemize}
	\item \textbf{Task}: Depending on the results of the reviewing the process of the initial data mining project, the project team decides how to proceed. The team decides whether (a) to continue to the deployment phase, (b) go back and refine or replace your models thus initiating further iterations, or (c) set up new data mining projects. This task includes analyses of remaining resources and budget, which may influence the decisions. If the results satisfactorily meet your data mining and business goals, start the deployment phase.

	\item \textbf{Outputs}: 
		\begin{itemize}
				\item {\color{red!50!black}{\textbf{List of possible actions}}}: List possible further actions along with the reasons for and against each option:  analyse potential for deployment and improvement(for each result obtained), recommend alternative following phases, refine the whole process, etc.
				\item {\textbf{Decision}}: Describe the decision made: rank alternatives, document reasons for the choice and  how to proceed along with the rationale.
 		\end{itemize}
	\end{itemize}
	
\begin{texample}
					\textbf{Next Steps} \par
					The bike rental company is fairly confident of both the accuracy and relevancy of the project results and so is continuing to the deployment phase.
\end{texample}

%%%%%%%%%%%%%%%%%%%%%%%%%%%%%%%%%%%%%%%%%%%%%%%%%%%%%%%%%%%%%%%%%%%%%%%%%%%%%%%%%%%%%%%%%%%%%%%%%%%%%%%%%%%%%%%%%%%%%%%%%%%%%%%%%%%%%%%%%%%%%%%%%%%%%%%%%%%%%%%%%%%%%%%%%%%%%%%%%%%%%%%%%%%%%%%%%%%%%%%%%%%%%%%%%%%%%%%%%%%%%%%%%%%%%%%%%%%%%%%%%%%%%%%%%%%%%%%%%%%%%%%%%%%%%%%%%%%%%%%%%%%%%%%%%%%%%%%%%%%%%%%%%%%%%%%%%%%%%%%%%%%%%%%%%%%%%%%%%%%%%%%%%%%%%%%%%%%%%%%%%%%%%%%%%%%%%%%%%%%%%%%%%%%%%%%%%%%%%%%%%%%%%%%%%%%%%%%%%%%%%%%%%%%%%%%%%%%%%%%%%%%%%%%%%%%%%%%%%%%%%%%%%%%%%%%%%%%%%%%%%%%%%%%%%%%%%%%%%%%%%%%%%%%%%%%%%%%%%%%%%%%%%%%%%%%%%%%%%%%%%%%%%%%%%%%%%%%%%%%%%%%%%%%%%%
\subsection{Deployment}

\begin{figure}[!ht]
	\centering
		\includegraphics[width=1.00\textwidth]{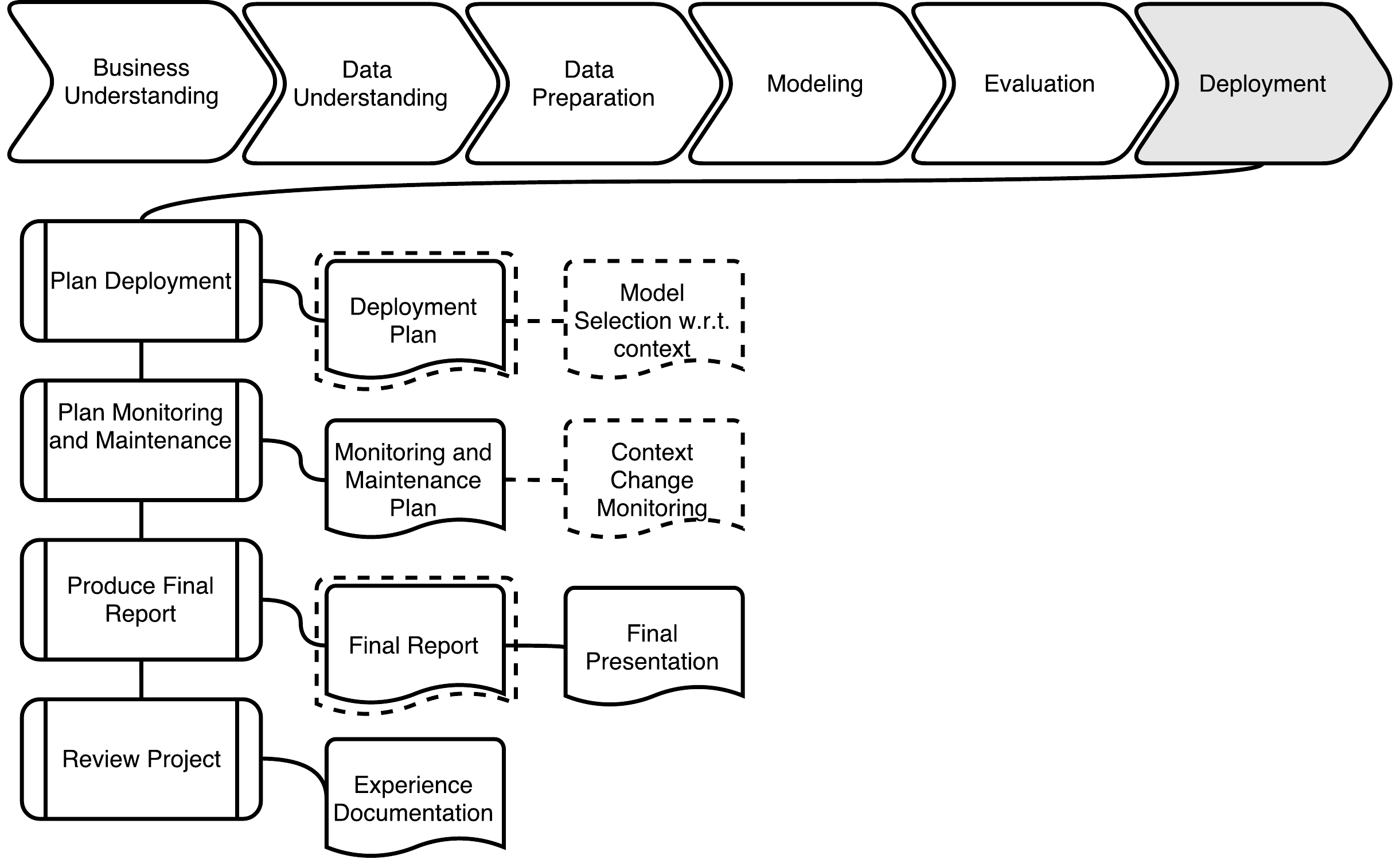}
	\caption{Phase 6. Deployment: tasks and activities for reframing.}
	\label{fig:Deployment}
\end{figure}

Creation of the model is generally not the end of the project and deployment is the process of using the discovered insights to make improvements (or changes) within your organization. Even if the results may not be formally integrated into your information systems, the knowledge gained will undoubtedly be useful for planning and making marketing decisions.This phase often involves planning and monitoring the deployment of results or completing wrap-up tasks such as producing a final report and conducting a project review.

Regarding context-awareness data mining tasks, in this phase we need to determine in what way the versatile model (or the pull of models) is to be kept, used, evaluated and maintained for a long-term use. Furthermore, we may need to monitor the possible change of the context distribution or check whether its range is the same as expected. If not, we may need to revaluate some models for a new distribution of contexts thus going back to previous phases/tasks.

\subsubsection{Plan deployment}
\begin{itemize}
	\item \textbf{Task}: Depending on the results of the reviewing the process of the initial data mining project, the project team decides how to proceed. The team decides whether (a) to continue to the deployment phase, (b) go back and refine or replace your models thus initiating further iterations, or (c) set up new data mining projects. This task includes analyses of remaining resources and budget, which may influence the decisions. If the results satisfactorily meet your data mining and business goals, start the deployment phase.

	\item \textbf{Outputs}: 
		\begin{itemize}
				\item {\color{red!50!black}{\textbf{Deployment plan}}}: This task takes the evaluation results and determines a strategy for deployment. If a general procedure has been identified to create the relevant model(s) and integrate within your database systems. This procedure is documented (step-by-step plan and integration) here for later deployment (including technical details, benefits of monitoring, deployment problems, etc.). Furthermore, create a plan to disseminate the relevant information to strategy makers
				\item {\color{red!50!black}{\textbf{Model selection w.r.t. the context}}}: Determine the pace in which context values are captured or estimated. Determine how the pool of models is going to be kept and selected according to context.
 		\end{itemize}
	\end{itemize}

\begin{texample}
					\textbf{Deployment Planning} \par
					Can we use different evaluation measures in modelling to achieve better results? How often do we need to retrain models?.
\end{texample}

\subsubsection{Plan deployment}
\begin{itemize}
	\item \textbf{Task}: Since your data mining work may be ongoing, monitoring and maintenance are important issues. In those cases the model(s) will likely need to be evaluated periodically to ensure its effectiveness and to make continuous improvements.

	\item \textbf{Outputs}: 
		\begin{itemize}
				\item {\textbf{Monitoring and maintenance plan}}: Summarize monitoring and maintenance strategy: factors or influences need to be tracked, validity and accuracy of each model, expiration issues,etc. 
				\item {\color{red!50!black}{\textbf{Context change Monitoring}}}: Determine the pace in which context values are captured or estimated. Determine how the pool of models is going to be kept and selected according to context.
 		\end{itemize}
	\end{itemize}

\begin{texample}
    			\textbf{Deployment Planning} \par
						The company had to decide on the criteria for model suitability for a given test station. In choosing the models to be reused in each of the new bike stations to be allocated, the bike rental company will select the set provided models (from existing stations) that have already been proved to be similar to the new station.	

\end{texample}

\subsubsection{Produce final report}
\begin{itemize}
	\item \textbf{Task}: At the end of the project, the project team writes up a final report to communicate the results

	\item \textbf{Outputs}: 
		\begin{itemize}
				\item {\color{red!50!black}{\textbf{Final report}}}: Final report where all the threads are brought together. It should include a thorough description of the original business problem, the process used to conduct data mining, how well initial data mining goals have been met, which (versatile) models are reused again and again,  budget and costs (cost of reframing? And retraining? How significant has context been?)), deviations from the original plan, summary of data mining results, overview of the deployment process, recommendations and insights discovered, etc.

				\item {\textbf{Final presentation}}: Determine the pace in which context values are captured or estimated. Determine how the pool of models is going to be kept and selected according to context.
 		\end{itemize}
	\end{itemize}

\subsubsection{Produce final report}
\begin{itemize}
	\item \textbf{Task}: This is the final step of the CASP-DM methodology. In it we assess what went right and what went wrong (and need to be improved), the final impressions, lessons learned, etc.

	\item \textbf{Outputs}: 
		\begin{itemize}
				\item {\color{red!50!black}{\textbf{Experience documentation}}}: Summarize important experiences made during the project (overall impressions, pitfalls, misleading approaches, etc.). Have contexts been well identified? How much model reuse has been performed? Were the models sufficiently flexible to be reframed? Should we change the definition of context? Can we make more versatile models?

 		\end{itemize}
	\end{itemize}

%%%%%%%%%%%%%%%%%%%%%%%%%%%%%%%%%%%%%%%%%%%%%%%%%%%%%%%%%%%%%%%%%%%%%%%%%%%%%%%%%%%%%%%%%%%%%%%%%%%%%%%%%%%%%%%%%%%%%%%%%%%%%%%%%%%%%%%%%%%%%%%%%%%%%%%%%%%%%%%%%%%%%%%%%%%%%%%%%%%%%%%%%%%%%%%%%%%%%%%%%%%%%%%%%%%%%%%%%%%%%%%%%%%%%%%%%%%%%%%%%%%%%%%%%%%%%%%%%%%%%%%%%%%%%%%%%%%%%%%%%%%%%%%%%%%%%%%%%%%%%%%%%%%%%%%%%%%%%%%%%%%%%%%%%%%%%%%%%%%%%%%%%%%%%%%%%%%%%%%%%%%%%%%%%%%%%%%%%%%%%%%%%%%%%%%%%%%%%%%%%%%%%%%%%%%%%%%%%%%%%%%%%%%%%%%%%%%%%%%%%%%%%%%%%%%%%%%%%%%%%%%%%%%%%%%%%%%%%%%%%%%%%%%%%%%%%%%%%%%%%%%%%%%%%%%%%%%%%%%%%%%%%%%%%%%%%%%%%%%%%%%%%%%%%%%%%%%%%%%%%%%%%%%%%%

%\input{3-Experimental.tex}
%\input{4-Previous.tex}
%\input{5-Conclusions.tex}
\section{Discussion}\label{lab:discussion}

Data mining is a discipline with strong technical roots in  statistics, machine learning and information systems. The advance in techniques, tools and platforms, jointly with the increase of the availability of data and the higher complexity of projects and teams, has been so significant in the past decade that methodological issues are becoming more important to harness all this potential in an efficient way. The perspective of data science, where data mining goals are more data-oriented than business-oriented in a more classical direct data mining process may suggest that rigid methodologies cannot cope with the variability of problems, which have to be adjusted to related scenarios very frequently, in terms of changes of data, goals, resolution, noise or utility functions. 

In contrast, we have advocated here that successful methodologies, such as CRISP-DM, can play this role if they become less rigid and accommodate the idea the variability of the application in a more systematic way. The notion of context, its identification and parametrisation, is a general way to anticipate all these changes and consider them from the very beginning. This is why CASP-DM tries to extend CRISP-DM to make this possible. The explicit existence of activity and tasks specifically designed for this context identification and handling ensures that companies and practitioners will not overlook this important aspect and will plan data mining projects in a more robust way, where data transformation and model construction can be reused and not jettisoned whenever any contextual thing changes. We have illustrated how CASP-DM goes through these context issues with some real examples.%\sidenoteJose{Last sentence. We haven't. Either section V improves or this sentence will have to go off.}

CASP-DM not only considers context-awareness in the whole process, but is backward compatible with CRISP-DM, the most common methodology in data mining. This means that CRISP-DM users can adopt CASP-DM immediately and even complement their existing projects with the context-aware bits, making them more versatile. In order to do this transition from CRISP-DM to CASP-DM, it is also important to have a stable platform and community where CASP-DM documents, phases and planning tools can be integrated and located for data mining practitioners. For instance, it is hard to find the CRISP-DM documentation, as nobody is maintaining it any more. To take that reference role, we have set up a community around \url{www.casp-dm.org}, where data mining practitioners can find information about CRISP-DM and CASP-DM, but also about context-awareness and other related areas such as reframing and domain adaptation.  It is also our intention to associate a working group with this initiative, so that CASP-DM can also evolve with the new methodological challenges of data mining.

%\section*{Acknowledgements}

\bibliographystyle{apalike}
\bibliography{biblio}

\begin{thebibliography}{}

\bibitem[Abowd et~al., 1999]{abowd1999towards}
Abowd, G.~D., Dey, A.~K., Brown, P.~J., Davies, N., Smith, M., and Steggles, P.
  (1999).
\newblock Towards a better understanding of context and context-awareness.
\newblock In {\em Handheld and ubiquitous computing}, pages 304--307. Springer.

\bibitem[Anand and B{\"u}chner, 1998]{anand1998decision}
Anand, S.~S. and B{\"u}chner, A.~G. (1998).
\newblock {\em Decision support using data mining}.
\newblock Financial Times Management.

\bibitem[Anand et~al., 1998]{anand1998data}
Anand, S.~S., Patrick, A., Hughes, J.~G., and Bell, D.~A. (1998).
\newblock A data mining methodology for cross-sales.
\newblock {\em Knowledge-Based Systems}, 10(7):449--461.

\bibitem[Angluin and Laird, 1988]{angluin1988learning}
Angluin, D. and Laird, P. (1988).
\newblock Learning from noisy examples.
\newblock {\em Machine Learning}, 2(4):343--370.

\bibitem[Bareinboim and Pearl, 2012]{bareinboim2012transportability}
Bareinboim, E. and Pearl, J. (2012).
\newblock Transportability of causal effects: Completeness results.
\newblock In {\em AAAI}.

\bibitem[Bi and Bennett, 2003]{bij2003regression}
Bi, J. and Bennett, K.~P. (2003).
\newblock Regression error characteristic curves.
\newblock In {\em Twentieth International Conference on Machine Learning
  (ICML-2003). Washington, DC}.

\bibitem[Blanco-Vega et~al., 2006]{blanco2006estimating}
Blanco-Vega, R., Ferri, C., Hern{\'a}ndez-Orallo, J., and
  Ram{\'\i}rez-Quintana, M.~J. (2006).
\newblock Estimating the class probability threshold without training data.
\newblock {\em ROC Analysis in Machine Learning}, page~9.

\bibitem[Brachman and Anand, 1996]{Brachman1996PKD}
Brachman, R.~J. and Anand, T. (1996).
\newblock Advances in knowledge discovery and data mining.
\newblock chapter The Process of Knowledge Discovery in Databases, pages
  37--57. American Association for Artificial Intelligence, Menlo Park, CA,
  USA.

\bibitem[Brunk et~al., 1997]{brunk1997mineset}
Brunk, C., Kelly, J., and Kohavi, R. (1997).
\newblock Mineset: An integrated system for data mining.
\newblock In {\em KDD}, pages 135--138.

\bibitem[Buchner et~al., 1999]{buchner1999internet}
Buchner, A.~G., Mulvenna, M.~D., Anand, S.~S., and Hughes, J.~G. (1999).
\newblock An internet-enabled knowledge discovery process.
\newblock In {\em Proceedings of the 9th international database conference,
  Hong Kong}, volume 1999, pages 13--27.

\bibitem[Cabena et~al., 1998]{cabena1998discovering}
Cabena, P., Hadjinian, P., Stadler, R., Verhees, J., and Zanasi, A. (1998).
\newblock {\em Discovering data mining: from concept to implementation}.
\newblock Prentice-Hall, Inc.

\bibitem[Caruana, 1997]{caruana1998multitask}
Caruana, R. (1997).
\newblock Multitask learning.
\newblock {\em Machine Learning}, 28(1):41--75.

\bibitem[Chapman et~al., 2000]{chapman2000crisp}
Chapman, P., Clinton, J., Kerber, R., Khabaza, T., Reinartz, T., Shearer, C.,
  and Wirth, R. (2000).
\newblock Crisp-dm 1.0 step-by-step data mining guide.

\bibitem[Chow, 1970]{chow1970optimum}
Chow, C. (1970).
\newblock On optimum recognition error and reject tradeoff.
\newblock {\em Information Theory, IEEE Transactions on}, 16(1):41--46.

\bibitem[Cios and Kurgan, 2005]{cios2005trends}
Cios, K.~J. and Kurgan, L.~A. (2005).
\newblock Trends in data mining and knowledge discovery.
\newblock In {\em Advanced techniques in knowledge discovery and data mining},
  pages 1--26. Springer.

\bibitem[Cios et~al., 2000]{cios2000knowledge}
Cios, K.~J., Teresinska, A., Konieczna, S., Potocka, J., and Sharma, S. (2000).
\newblock A knowledge discovery approach to diagnosing myocardial perfusion.
\newblock {\em Engineering in Medicine and Biology Magazine, IEEE},
  19(4):17--25.

\bibitem[Debuse et~al., 2001]{debuse2001building}
Debuse, J., de~la Iglesia, B., Howard, C., and Rayward-Smith, V. (2001).
\newblock Building the kdd roadmap.
\newblock In {\em Industrial Knowledge Management}, pages 179--196. Springer.

\bibitem[Drummond and Holte, 2006]{drummond-and-Holte2006}
Drummond, C. and Holte, R. (2006).
\newblock {Cost Curves: An Improved Method for Visualizing Classifier
  Performance}.
\newblock {\em Machine Learning}, 65:95--130.

\bibitem[Edelstein, 1998]{edelstein1998introduction}
Edelstein, H.~A. (1998).
\newblock {\em Introduction to data mining and knowledge discovery}.
\newblock Two Crows.

\bibitem[Elkan, 2001]{Elk01}
Elkan, C. (2001).
\newblock The foundations of {Cost-Sensitive} learning.
\newblock In {\em {IJCAI}-01}, pages 973--978.

\bibitem[Fawcett, 2006]{fawcett2006introduction}
Fawcett, T. (2006).
\newblock An introduction to {ROC} analysis.
\newblock {\em Pattern recognition letters}, 27(8):861--874.

\bibitem[Fayyad et~al., 1996a]{Fayyad1996KDD}
Fayyad, U., Piatetsky-Shapiro, G., and Smyth, P. (1996a).
\newblock The kdd process for extracting useful knowledge from volumes of data.
\newblock {\em Commun. ACM}, 39(11):27--34.

\bibitem[Fayyad et~al., 1996b]{fayyad1996advances}
Fayyad, U.~M., Piatetsky-Shapiro, G., Smyth, P., and Uthurusamy, R. (1996b).
\newblock Advances in knowledge discovery and data mining.

\bibitem[Ferri et~al., 2009]{PRL09}
Ferri, C., Hern\'{a}ndez-Orallo, J., and Modroiu, R. (2009).
\newblock An experimental comparison of performance measures for
  classification.
\newblock {\em Pattern Recognition Letters}, 30(1):27--38.

\bibitem[Flach, 2010]{flach2010roc}
Flach, P. (2010).
\newblock {ROC} analysis.
\newblock In {\em Encyclopedia of Machine Learning}, pages 869--875. Springer.

\bibitem[Flach et~al., 2003]{flach2003decision}
Flach, P., Blockeel, H., Ferri, C., Hern{\'a}ndez-Orallo, J., and Struyf, J.
  (2003).
\newblock Decision support for data mining.
\newblock In {\em Data Mining and Decision Support}, pages 81--90. Springer.

\bibitem[Flach et~al., 2011]{ICML11CoherentAUC}
Flach, P., Hern{\'a}ndez-Orallo, J., and Ferri, C. (2011).
\newblock A coherent interpretation of {AUC} as a measure of aggregated
  classification performance.
\newblock In {\em ICML}.

\bibitem[Fr{\'e}nay and Verleysen, 2013]{frenay2013classification}
Fr{\'e}nay, B. and Verleysen, M. (2013).
\newblock Classification in the presence of label noise: a survey.
\newblock {\em IEEE Transactions on Neural Networks and Learning Systems},
  25(5).

\bibitem[Gertosio and Dussauchoy, 2004]{gertosio2004knowledge}
Gertosio, C. and Dussauchoy, A. (2004).
\newblock Knowledge discovery from industrial databases.
\newblock {\em Journal of Intelligent Manufacturing}, 15(1):29--37.

\bibitem[Giraud-Carrier et~al., 2004]{giraud2004introduction}
Giraud-Carrier, C., Vilalta, R., and Brazdil, P. (2004).
\newblock Introduction to the special issue on meta-learning.
\newblock {\em Machine learning}, 54(3):187--193.

\bibitem[Hand, 2009]{hand2009measuring}
Hand, D. (2009).
\newblock {Measuring classifier performance: a coherent alternative to the area
  under the ROC curve}.
\newblock {\em Machine learning}, 77(1):103--123.

\bibitem[Harry, 1998]{harry1998six}
Harry, M.~J. (1998).
\newblock Six sigma: a breakthrough strategy for profitability.
\newblock {\em Quality progress}, 31(5):60.

\bibitem[Hern\'andez-Orallo, 2013]{RROC2013}
Hern\'andez-Orallo, J. (2013).
\newblock {ROC} curves for regression.
\newblock {\em Pattern Recognition}, 46(12):3395--3411.

\bibitem[Hern{\'a}ndez-Orallo et~al., 2015]{Hernandez-Orallo2015}
Hern{\'a}ndez-Orallo, J., Ferri, C., Lachiche, N., Mart{\'i}nez-Us{\'o}, A.,
  and Ram{\'i}rez-Quintana, M.~J. (2015).
\newblock Binarised regression tasks: methods and evaluation metrics.
\newblock {\em Data Mining and Knowledge Discovery}, pages 1--43.

\bibitem[Hern{\'a}ndez-Orallo et~al., 2016]{Orallo2016DMKD}
Hern{\'a}ndez-Orallo, J., Ferri, C., Lachiche, N., Mart{\'i}nez-Us{\'o}, A.,
  and Ram{\'i}rez-Quintana, M.~J. (2016).
\newblock Binarised regression tasks: methods and evaluation metrics.
\newblock {\em Data Mining and Knowledge Discovery}, 30(4):848--890.

\bibitem[Hern\'{a}ndez-Orallo et~al., 2011]{ICML11Brier}
Hern\'{a}ndez-Orallo, J., Flach, P., and Ferri, C. (2011).
\newblock Brier curves: a new cost-based visualisation of classifier
  performance.
\newblock In {\em ICML}.

\bibitem[Hern{\'a}ndez-Orallo et~al., 2012a]{hernandez2012unified}
Hern{\'a}ndez-Orallo, J., Flach, P., and Ferri, C. (2012a).
\newblock A unified view of performance metrics: Translating threshold choice
  into expected classification loss.
\newblock {\em JMLR}, 13:2813--2869.

\bibitem[Hern{\'a}ndez-Orallo et~al., 2012b]{JMLR2012}
Hern{\'a}ndez-Orallo, J., Flach, P., and Ferri, C. (2012b).
\newblock A unified view of performance metrics: Translating threshold choice
  into expected classification loss.
\newblock {\em Journal of Machine Learning Research}, 13:2813--2869.

\bibitem[Hern\'andez-Orallo et~al., 2013]{MLJ2013}
Hern\'andez-Orallo, J., Flach, P., and Ferri, C. (2013).
\newblock {ROC} curves in cost space.
\newblock {\em Machine Learning}, 93(1):71--91.

\bibitem[Hern{\'{a}}ndez{-}Orallo et~al., 2016]{aicomHernandez16}
Hern{\'{a}}ndez{-}Orallo, J., Us{\'{o}}, A.~M., Prud{\^{e}}ncio, R. B.~C.,
  Kull, M., Flach, P.~A., Ahmed, C.~F., and Lachiche, N. (2016).
\newblock Reframing in context: {A} systematic approach for model reuse in
  machine learning.
\newblock {\em {AI} Commun.}, 29(5):551--566.

\bibitem[Jiang, 2008]{jiang2008literature}
Jiang, J. (2008).
\newblock A literature survey on domain adaptation of statistical classifiers.
\newblock {\em URL: http://sifaka. cs. uiuc.
  edu/jiang4/domainadaptation/survey}.

\bibitem[Khreich et~al., 2012]{Khreich:IS12}
Khreich, W., Granger, E., Miri, A., and Sabourin, R. (2012).
\newblock A survey of techniques for incremental learning of {HMM} parameters.
\newblock {\em Information Sciences}, 197:105--130.

\bibitem[Kull and Flach, 2014]{datashiftpatterns2014}
Kull, M. and Flach, P. (2014).
\newblock Patterns of dataset shift.
\newblock In {\em Ws. on Learning over Multiple Contexts at ECML2014 (LMCE)}.

\bibitem[Kull and Hern\'andez-Orallo, 2015]{MissingAttributes2014}
Kull, M. and Hern\'andez-Orallo, J. (2015).
\newblock Missing values on purpose: Model selection and reframing with
  attribute and prediction costs.
\newblock {\em submitted}.

\bibitem[Kull et~al., 2015a]{kullmorebikes}
Kull, M., Lachiche, N., and Mart{\i}nez-Us{\'o}, A. (2015a).
\newblock Morebikes-model reuse with bike rental station data.

\bibitem[Kull et~al., 2015b]{KullLU15}
Kull, M., Lachiche, N., and Us{\'{o}}, A.~M. (2015b).
\newblock Model reuse with bike rental station data (preamble).
\newblock In {\em Proceedings of the {ECML/PKDD} 2015 Discovery Challenges
  co-located with European Conference on Machine Learning and Principles and
  Practice of Knowledge Discovery in Databases {(ECML-PKDD} 2015), Porto,
  Portugal, September 7-11, 2015.}

\bibitem[Lo et~al., 2011]{Lo:Wang:Wang:Lin:TransOnM11}
Lo, H.-Y., Wang, J.-C., Wang, H.-M., and Lin, S.-D. (2011).
\newblock Cost-sensitive multi-label learning for audio tag annotation and
  retrieval.
\newblock {\em Multimedia, IEEE Transactions on}, 13(3):518--529.

\bibitem[Mariscal et~al., 2010]{mariscal2010survey}
Mariscal, G., Marban, O., and Fernandez, C. (2010).
\newblock A survey of data mining and knowledge discovery process models and
  methodologies.
\newblock {\em The Knowledge Engineering Review}, 25(02):137--166.

\bibitem[Mart\'{\i}nez-Us\'o and Hern\'andez-Orallo, 2015]{multidimensional}
Mart\'{\i}nez-Us\'o, A. and Hern\'andez-Orallo, J. (2015).
\newblock Multidimensional prediction models when the resolution context
  changes.
\newblock In {\em ECML}.

\bibitem[Mart{\'i}nez-Us{\'o} et~al., 2015]{AEPIA15}
Mart{\'i}nez-Us{\'o}, A., Hern{\'a}ndez-Orallo, J., Ram{\'i}rez-Quintana,
  M.~J., and Plumed, F.~M. (2015).
\newblock {\em Pentaho + R: An Integral View for Multidimensional Prediction
  Models}, pages 234--244.
\newblock Springer International Publishing.

\bibitem[Metz, 1978]{metz1978basic}
Metz, C.~E. (1978).
\newblock Basic principles of {ROC} analysis.
\newblock In {\em Seminars in nuclear medicine}, volume 8,4, pages 283--298.
  Elsevier.

\bibitem[Moreno-Torres et~al., 2012]{moreno2012unifying}
Moreno-Torres, J.~G., Raeder, T., Alaiz-Rodr{\'\i}guez, R., Chawla, N.~V., and
  Herrera, F. (2012).
\newblock A unifying view on dataset shift in classification.
\newblock {\em Pattern Recognition}, 45(1):521--530.

\bibitem[Moyle and Jorge, 2001]{moyle2001ramsys}
Moyle, S. and Jorge, A. (2001).
\newblock Ramsys-a methodology for supporting rapid remote collaborative data
  mining projects.
\newblock In {\em ECML/PKDD 2001 Workshop on Integrating Aspects of Data
  Mining, Decision Support and Meta-Learning: Internal SolEuNet Session}, pages
  20--31.

\bibitem[Pan and Yang, 2010]{pan2010survey}
Pan, S.~J. and Yang, Q. (2010).
\newblock A survey on transfer learning.
\newblock {\em Knowledge and Data Engineering, IEEE Transactions on},
  22(10):1345--1359.

\bibitem[Pietraszek, 2007]{pietraszek2007use}
Pietraszek, T. (2007).
\newblock On the use of {ROC} analysis for the optimization of abstaining
  classifiers.
\newblock {\em Machine Learning}, 68(2):137--169.

\bibitem[Qui{\~n}onero-Candela et~al., 2009]{quinonero2009}
Qui{\~n}onero-Candela, J., Sugiyama, M., Schwaighofer, A., and Lawrence, N.~D.
  (2009).
\newblock {\em Dataset shift in machine learning}.
\newblock The MIT Press.

\bibitem[Raedt, 1992]{DRa92}
Raedt, L.~D. (1992).
\newblock {\em Interactive Theory Revision: An Inductive Logic Programming
  Approach}.
\newblock Academic Press.

\bibitem[Richards and Mooney, 1991]{richards1991first}
Richards, B.~L. and Mooney, R.~J. (1991).
\newblock First-order theory revision.
\newblock In {\em ML}, pages 447--451.

\bibitem[SAS, 2005]{SEMMA}
SAS (2005).
\newblock Semma data mining methodology.
\newblock
  \url{http://www.sas.com/technologies/analytics/datamining/miner/semma.html}.

\bibitem[Scheirer et~al., 2013]{scheirer2013toward}
Scheirer, W.~J., de~Rezende-Rocha, A., Sapkota, A., and Boult, T.~E. (2013).
\newblock Toward open set recognition.
\newblock {\em Pattern Analysis and Machine Intelligence, IEEE Transactions
  on}, 35(7):1757--1772.

\bibitem[Thrun, 1996]{thrun1996learning}
Thrun, S. (1996).
\newblock Is learning the n-th thing any easier than learning the first?
\newblock {\em Advances in neural information processing systems}, pages
  640--646.

\bibitem[Thrun and Pratt, 2012]{thrun2012learning}
Thrun, S. and Pratt, L. (2012).
\newblock {\em Learning to learn}.
\newblock Springer Science \& Business Media.

\bibitem[Torrey and Shavlik, 2009]{torrey2009transfer}
Torrey, L. and Shavlik, J. (2009).
\newblock Transfer learning.
\newblock {\em Handbook of Research on Machine Learning Applications},
  3:17--35.

\bibitem[Tortorella, 2005]{tortorella2005roc}
Tortorella, F. (2005).
\newblock A {ROC}-based reject rule for dichotomizers.
\newblock {\em Pattern Recognition Letters}, 26(2):167--180.

\bibitem[Turney, 2000]{turney2000types}
Turney, P. (2000).
\newblock Types of cost in inductive concept learning.
\newblock {\em Canada National Research Council Publications Archive}.

\bibitem[Vanderlooy et~al., 2006]{vanderlooy2006analysis}
Vanderlooy, S., Sprinkhuizen-Kuyper, I., and Smirnov, E. (2006).
\newblock An analysis of reliable classifiers through {ROC} isometrics.
\newblock In {\em Proceedings of the ICML 2006 Ws. on {ROC} Analysis (ROCML
  2006), Pittsburgh, USA, June}, volume~29, pages 55--62.

\bibitem[Xu et~al., 2014]{Xu:Kusner:Weinberger:Chen:Chapelle:JMLR14}
Xu, Z., Kusner, M.~J., Weinberger, K.~Q., Chen, M., and Chapelle, O. (2014).
\newblock Classifier cascades and trees for minimizing feature evaluation cost.
\newblock {\em JMLR}, 15:2113--2144.

\end{thebibliography}

\end{document}